\documentclass[letterpaper,11pt]{article}
\usepackage[utf8]{inputenc}

\usepackage{jheppub}
\usepackage{graphicx,verbatim}
\usepackage{blindtext}
\usepackage[T1]{fontenc}
\usepackage{epsf}
\usepackage{lmodern}
\usepackage[mathscr]{euscript}

\usepackage{ytableau}
\usepackage{youngtab}

\usepackage{amsmath}
\usepackage{amsfonts}
\usepackage{amssymb}
\usepackage{mathrsfs}
\usepackage{slashed}
\usepackage{xcolor}
\usepackage{tikz, float}
\usetikzlibrary{patterns,shapes.misc}

\def\kq{\mathfrak{q}}

\def\ri{\mathrm{i}}
\def\rj{\mathrm{j}}

\def\bz{\mathbf{z}}

\def\bx{\mathbf{x}}
\def\ba{\mathbf{a}}
\def\bb{\mathbf{b}}

\def\bT{\mathbf{T}}

\def\bX{\mathbf{X}}

\def\bM{\mathbf{M}}

\def\bJ{\mathbf{J}}
\def\bK{\mathbf{K}}
\def\bN{\mathbf{N}}
\def\bS{\mathbf{S}}

\def\BC{\mathbb{C}}

\def\BR{\mathbb{R}}
\def\BE{\mathbb{E}}
\def\BZ{\mathbb{Z}}
\def\BI{\mathbb{I}}
\def\BJ{\mathbb{J}}

\def\CalN{\mathcal{N}}

\def\CalR{\mathcal{R}}
\def\CalZ{\mathcal{Z}}

\def\CalO{\mathcal{O}}

\def\Tr{{\rm Tr}}

\def\ve{{\varepsilon}}

\def\tg{\mathtt{g}}
\def\tp{\mathtt{p}}
\def\tq{\mathtt{q}}

\def\fM{\mathfrak{M}}

\def\sT{\mathsf{T}}

\def\sn{\mathsf{n}}
\def\sx{\mathsf{x}}

\def\sH{\mathsf{H}}

\def\EX{\EuScript{X}}

\def\EH{\EuScript{H}}

\def\EK{\EuScript{K}}
\def\EW{\EuScript{W}}

 \def\p{\partial}
 
 \def\a{\alpha}
 \def\b{\beta}
 \def\g{\gamma}
 \def\d{\delta}

 \def\G{\Gamma}

 \def\o{\omega }

  \def\sh{\text{sh}}

\def\beq{\begin{equation}}
\def\eeq{\end{equation}}

\title{Defects and type D relativistic Toda lattice for some 5d gauge theories}
\author[a,b]{Kimyeong Lee}
\author[c]{and Norton Lee}
\affiliation[a]{School of Physics, Korea Institute for Advanced Study,\\Hoegiro 85, Seoul 02455, Korea}
\affiliation[b]{ 
Beijing Institute of Mathematical Sciences and Applications (BIMSA), \\
Huaibei Town, Huairou District, Beijing 101408, China}

\affiliation[c]{Center for Geometry and Physics, Institute for Basic Science (IBS), \\Pohang 37673, Korea}
\emailAdd{klee@kias.re.kr, norton.lee@ibs.re.kr}

\preprint{CGP24011}

\vspace{2cm}
\abstract{We perform folding on the ADHM construction of the instanton moduli space from $SU$ to $SO$ group. A Young diagram description for the $SO$ instanton is obtained after modifying the real and complex moment maps of the ADHM data. We study the Bethe gauge correspondence between type D relativistic Toda lattice and 5d $\CalN=1$ folded theory. In particular we prove that the regular monodromy defect in the folded gauge theory is the stationary wavefunction of the type D relativistic Toda lattice. }

\begin{document}

\maketitle


\newpage

\section{Introduction}

Supersymmetric gauge theories with eight supercharges have intrinsic integrability, known as the \emph{Bethe/Gauge correspondence}. It has been an active research area since the ground breaking work by Seiberg and Witten \cite{Seiberg:1994aj,Seiberg:1994rs}. 
The R-matrix characterizing the quantum integrability is found associated to the instanton counting of gauge theory with an $\Omega$-background \cite{maulik2012quantum,nakajima2013quiver}. In the Nekrasov-Shatashvili limit where 2d $\CalN=(2,2)$ super-Poincar\'{e} symmetry is restored the 4d $\CalN=2$ theory is effectively described by an effective $\CalN=(2,2)$ theory that captures the integrability nature \cite{Nikita-Shatashvili-1,Nikita-Shatashvili-2,Nikita-Shatashvili-3}. 
When the gauge group of the 4d gauge theory is $G=A_N$, the R-matrix is associated to a coproduct of $N$ copies of affine Yangian of $\mathfrak{gl}_1$ \cite{Jeong:2024hwf,Jeong:2024mxr}. In particular, the R-matrices of the integrabile system can be explicitly constructed from the 4d gauge theory. 
The stationary state of the quantum integrable system -- under Bethe/Gauge correspondence -- are the vacuua of the effective 2d $\CalN=(2,2)$ theory. 
This beautiful correspondence has a K-theoretic uplift: The R-matrix, associated with the Ding-Iohara-Miki algebra \cite{ding1996generalization,miki2007q,feigin2011quantum}, can be constructed from 5d $\CalN=1$ gauge theory with gauge group $G = A_N$ \cite{Lee:2023wbf}. 
The stationary state of the quantum integrable system are the vacuua of the effective 3d $\CalN=2$ theory. 

When the gauge group $G$ is of BCD-type, the action of the quantum toroidal algebra on the supersymmetric gauge theory is less well understood. Unlike A-type, the instanton configurations of gauge theories with BCD-type gauge groups lack a Young diagram description. The $qq$-character -- which acts as a quantum-quantum uplift of the Seiberg-Witten curve -- for BCD-type \cite{Haouzi:2020yxy} is much more complicated than for A-type \cite{Nikita:I,Kimura:2015rgi,Bourgine:2015szm,Bourgine:2016vsq} in the general $\Omega$-background\footnote{Young diagram descriptions of instanton configurations exist in the unrefined limit for BCD-type \cite{Nawata:2021dlk,Nawata:2023wnk}. However such limit does not yield quantum integrability.}.

\paragraph{}
In this work, our goal is to study the Bethe/Gauge correspondence between the quantum type D relativistic Toda lattice (RTL for short) and 5d $\CalN=1$ supersymmetric gauge theory. 
On the integrable systems side, RTL of type BCD can be treated as type A with specific boundary condition \cite{sklyanin1988boundary}.  
On the gauge theory side, 
We fold the $SU(2N)$ gauge theory with eight fundamental matters into the $SO(2N)$ gauge theory by fine-tuning the Coulomb moduli parameters and the masses of the hypermultiplets. By restricting the $SU(2N)$ ADHM moduli space to the $SO$ sector, with the real and complex moment maps properly modified, we obtain a Young diagram description for the $SO$ gauge theory. 
The instanton partition function is then given by an ensemble of Young diagrams similar to type A, with a much simpler $qq$-character associated to.
The effective potential of 3d $\CalN=2$ theory can be obtained from the folded partition function in the Nekrasov-Shatashvili limit, whose vacuum equation is identified with the Bethe ansatz equation of the type D RTL. Alternatively, the same equation can be obtained by the analycity of the $qq$-character \cite{Chen:2019vvt,Chen:2020rxu}. 

To find the stationary wavefunction of type D RTL, we introduce the regular surface defect \cite{nekrasov20042d}, also known as the Gukov-Witten monodromy-type defect \cite{Gukov:2008sn,Gukov:2006jk}, defined by a singular boundary condition along a surface, which can be modeled by the orbifold construction. The parameters of the co-dimensional two defect become the coordinates on which the stationary states depend.
The defect partition function obeys non-perturbative Dyson-Schwinger equation, which becomes Schr\"{o}dinger-type equation in the Nekrasov-Shatashvili limit \cite{Nikita:V}. 

\subsection*{The main results and outline}

The main results of this paper are: 
\begin{itemize}
    \item Establishing the folding from type A ADHM construction to type D, resulting Young diagram descriptions of instanton configuration for $SO(2N)$ gauge group. 
    \item The Bethe ansatz equation of the type D RTL is recovered from the gauge theory through both the vacuum equation of the twisted superpotential and analyticity of $qq$-character. 
    \item The identification of the  vacuum expectation value of the monodromy defect $\psi$ being the stationary state of the type D RTL. This is done by exploiting the analyticity property of the $qq$-character in the presence of monodromy defect.
\end{itemize}

This article is organized as follows. In Section.~\ref{sec:RTL} we give a brief review on the integrability of relativistic Toda lattice. 
In Section.~\ref{sec:partition function} we review on the ADHM construction for $SU$ and $SO$ gague group instanton. We fold the instanton partition function from $SU$ to $SO$ gauge group. 
By using the folded partition function, we first reproduce the well-known Bethe/Gauge correspondence between the
Bethe ansatz equation of type $D$ relativistic Toda lattice and the $qq$-character in the Nekrasov-Shatashvili limit in Section.~\ref{sec:gauge}. 
We further push forward by introducing co-dimensional two monodromy defect to the gauge theory. By exploiting the analytic property of fractional $qq$-character, we reproduce the quantum Hamiltonian of type $D$ RTL from the gauge theory and identify the defect partition function as its wavefunction.
Finally we point out our conclusion and point out furture direction in Section.~\ref{sec:summary}.
 
\acknowledgments

The authors thank Saebyeok Jeong, Taro Kimura, Yongchao L\"{u}, Xin Wang for the discussion. 
The work of NL is supported by IBS project grant IBS-R003-D1.  
KL is supported in part by KIAS Grants PG006904 and the National Re-
search Foundation of Korea (NRF) Grant funded by the Korea government (MSIT) (No.
2017R1D1A1B06034369). KL also thanks KITP for the program \href{https://www.kitp.ucsb.edu/activities/strings24}{What is
String Theory? Weaving Perspectives Together}. This research was supported in part by grant NSF PHY-
2309135 to the Kavli Institute for Theoretical Physics (KITP). 
We would like to thank SCGP for the \href{https://scgp.stonybrook.edu/archives/41264}{21st Simons Physics Summer Workshop "Landscapia"} where the part of work was done.

\section{Relativistic Toda lattice}\label{sec:RTL}

The integrability of Relativistic Toda lattice (RTL for short) is characterized by the $R$-matrix $R_{a_i,a_j} :V_{a_i} \otimes V_{a_j} \to V_{a_i} \otimes V_{a_j}$ , satisfying the Yang-Baxter equation
\begin{align}
    R_{a_1,a_2}(x-x') R_{a_1,a_3}(x) R_{a_2,a_3}(x') = R_{a_2,a_3}(x') R_{a_1,a_3}(x) R_{a_1,a_2}(x-x').
\end{align}
The $2\times 2$ Lax matrix is a special case of the $R$-matrix with the choice of 
\[
    V_{a_1}=V_{a_2} = \BC^2 := V_{\rm aux}, \ V_{a_3} = \EH_n
\]
on each lattice site.
$\EH_n$ is the Hilbert space of the $n$-th particle. $V_{\rm aux}$ is called the \emph{auxiliary space}. 
On each of the lattice site we define a $2 \times 2$ Lax operator as a $GL_2$-valued function  \cite{Kuznetsov:1994ur,Iorgov:2007ks,deVega:1993xi}:
\begin{align}
    L_n(x) = \begin{pmatrix}
        \sh (x - \hat\tp_n) & -R e^{-\tq_n} \\
         R e^{\tq_n} & 0
    \end{pmatrix}\in \text{End}(\EH_n \otimes V_\text{aux})
\end{align}
where $\hat\tp_n=\hbar \p_{\tq_n}$ and $\tq_n$ are the canonically conjugated momentum and coordinate of the $n$-th particle. $R$ is the potential constant.
For later convenience, we define
\[
    \sh(x) = 2\sinh \left(\frac{x}{2} \right).
\]
The R-matrix acting on the space $V_\text{aux} \otimes V_\text{aux}$ by
\begin{align}\label{def:R-matrix}
    R_{a_1,a_2}(x-x') = 
    \begin{pmatrix} 
        \sh(x-x'+\hbar) & 0 & 0 & 0 \\
        0 & \sh (x-x') & \sh\hbar & 0 \\
        0 & \sh\hbar & \sh(x-x') & 0 \\
        0 & 0 & 0 & \sh(x-x'+\hbar)
    \end{pmatrix} . 
\end{align}

The commutation relations between two elements in the Lax operator is governed by the Yang-Baxter RLL-relation (train track relation)
\begin{align}\label{eq:RLL}
    R_{a_1,a_2}(x-x')L_{a_1}(x) L_{a_2}(x') = L_{a_2}(x') L_{a_1}(x) R_{a_1,a_2}(x-x')
\end{align}
which can be verified true by direct computation.  


The monodromy matrix $\bT(x)$ of type A RTL with periodic boundary condition is an ordered product of the Lax matrices across $N$ particles
\begin{align}
    \bT(x) = L_N(x) L_{N-1}(x) \cdots L_2(x) L_1(x) \in \text{End} \left(\bigotimes_{n=1}^N \EH \otimes V_{\rm aux} \right).
\end{align}
It is obvious that the monodromy matrix $\bT(x)$ satisfies the same Yang-Baxter equation as the Lax operator: 
\begin{align}\label{eq:RTT}
    R_{a_1,a_2}(x-x')\bT_{a_1}(x) \bT_{a_2}(x') = \bT_{a_2}(x') \bT_{a_1}(x) R_{a_1,a_2}(x-x').
\end{align}
The spectral curve of the integrable system is defined through introduction of spectral parameter $y$:
\begin{align}
    \text{q-det} (\bT(x) - y) = y^2 - \Tr \bT(x) y + \text{q-det}\bT(x) = 0. 
\end{align}

\subsection{Toda lattice with boundary}

E. Sklyanin points out that the monodromy matrix of type BCD RTL can be obtained by introducing reflection matrix as boundary condition \cite{sklyanin1988boundary}.
The transfer matrix of a RTL with boundary condition is given by
\begin{align}
    t(x) = \Tr K_+(x) \bT(x) K_-(x) \bT^{-1}(-x). 
\end{align}
$\bT(x)$ is usually taken as the same as closed RTL.  
$K_\pm(x)$ are the reflection matrices obeying the reflection equation
\begin{align}
\begin{split}
    & R_{12}(x-x') K_{+,1}(x) R_{21}(x+x'-\hbar) K_{+,2}(x') \\ 
    & = K_{+,2}(x') R_{12}(x+x'-\hbar) K_{+,1}(x) R_{21}(x-x')
\end{split}
\end{align}
and 
\begin{align}
\begin{split}
    & R_{12}(-x+x') K^{T}_{-,1}(x) R_{21}(-x-x'-\hbar) K^{T}_{-,2}(x') \\
    & = K^{T}_{-,2}(x') R_{12}(-x-x'-\hbar) K^{T}_{-,1}(x) R_{21}(-x+x').
\end{split}
\end{align}
The reflection matrices for type BC RTL are constant matrices $K_{\pm} \in \text{End}(V_\text{aux})$, $\bT(x) \in \text{End}(\bigotimes^N_{n=1} \EH_n \otimes V_\text{aux})$. 
In the case for type D, the reflection matrices depends on dynamical variables $K_+ \in \text{End}(\EH_1 \otimes V_\text{aux})$, $K_- \in \text{End}(\EH_N \otimes V_\text{aux})$, $\bT(x) \in \text{End}(\bigotimes^{N-1}_{n=2} \EH_n \otimes V_\text{aux})$. 

Given a simple solution $K_\pm(x)$ to the reflection equation, one can check that
\begin{subequations}
\begin{align}
    U^{T}_+(x) & =  \bT_+(x) K_+(x) (\bT^{-1}_+(-x))^T,  \\
    U_-(x) & = \bT_-(x) K_-(x) \bT_-^{-1}(-x)  
\end{align}
\end{subequations}
satisfy the same reflection equation if $\bT_\pm(x)$ satisfy \eqref{eq:RTT}. 
The generating function of the integral of motion is 
\begin{align}
\begin{split}
    t(x) & = \Tr U_+ U_- = \Tr K_+(x) \bT(x) K_-(x) \bT(-x)^{-1} \\
    & = e^{Nx} \left[ 1 + \sum_{n=1}^N H_n e^{-2nx}  \right].
\end{split}
\end{align}
The reflection matrix $K_\pm$ satisfying the reflection equation takes the following form:
\begin{subequations}
\begin{align}
    K_+ & = \begin{pmatrix}
        \alpha_1^+ e^{\frac{x}{2}} - \alpha_2^+ e^{-\frac{x}{2}} & \d^+(e^x + e^{-x}) - \beta^+ \\
        \g^+ - \d^+(e^x + e^{-x}) & \alpha_2^+ e^{\frac{x}{2}} - \alpha_1^+ e^{-\frac{x}{2}}
    \end{pmatrix}, \\
    K_- & = \begin{pmatrix}
        - \alpha_1^- e^{\frac{x}{2}} + \alpha_2^- e^{-\frac{x}{2}} & \g^- - \d^-(e^x + e^{-x}) \\
        \d^- (e^x + e^{-x} ) - \b^- & -\a_2^- e^{\frac{x}{2}} + \a_1^- e^{-\frac{x}{2}}
    \end{pmatrix}.
\end{align}
\end{subequations}
The first Hamiltonian obtained from the transfer matrix reads 
\begin{align}
\begin{split}
    H_{1} = & \sum_{j=1+\chi}^{N-\chi} 2\cosh \tp_j + \sum_{j=1+\chi}^{N-1-\chi} R^2 e^{\tq_j-\tq_{j+1}} 2\cosh \frac{\tp_j+\tp_{j+1}}{2} \\
    & + \b^+ + \b^- + R \alpha_1^+ e^{-\frac{\tp_{1+\chi}}{2}-\tq_{1+\chi}} + R \alpha_2^+ e^{\frac{\tp_{1+\chi}}{2}-\tq_{1+\chi}} + R \alpha_1^- e^{-\frac{\tp_{N-\chi}}{2}+\tq_{N-\chi}} + R \a_2^- e^{\frac{\tp_{N-\chi}}{2}+\tq_{N-\chi}} \\
    & + \d^+ R^2  e^{-2\tq_{1+\chi}} + \d^- R^2 e^{2\tq_{N-\chi}} \\
\end{split}
\end{align}
with
\[
    \chi = \begin{cases}
        0, & \text{type B \& C,} \\
        1, & \text{type D.}
    \end{cases}
\]
The boundary reflection matrices $K_\pm$ of type D RTL are \cite{Kuznetsov:1994ur}:
\begin{subequations}\label{eq:twistK-D}
\begin{align}
    K_+ & = 
    \begin{pmatrix}
        R [e^{\frac{x}{2}} \cosh(\tp_1/2-\tq_1) - e^{-\frac{x}{2}} \cosh(\tp_1/2+\tq_1) ] & \cosh x  - \cosh {\tp_1} \\
        2R^2 \cosh 2\tq_1 - 2\tg^2 \cosh {x} & R [-e^{\frac{x}{2}} \cosh(\tp_1/2+\tq_1) + e^{-\frac{x}{2}} \cosh(\tp_1/2-\tq_1) ] 
    \end{pmatrix}, \\
    K_- & = 
    \begin{pmatrix}
        R [ -e^{\frac{x}{2}} \cosh(\tp_N/2+\tq_N) + e^{-\frac{x}{2}}\cosh(\tp_N/2-\tq_N)] & 2R^2 \left[\cosh 2\tq_N - \cosh x \right] \\ 
        2\cosh x - 2\cosh {\tp_{N}} & R [e^{\frac{x}{2}} \cosh(\tp_N/2-\tq_N) - e^{-\frac{x}{2}} \cosh(\tp_N/2+\tq_N)]
    \end{pmatrix}.
\end{align}
\end{subequations}
The first Hamiltonian of $\hat{D}_N$ RTL is
\begin{align}\label{def:H-D-RTL}
\begin{split}
    H_{1} = & \sum_{j=1}^{N} 2\cosh \tp_j + \sum_{j=1}^{N-1} R^2 e^{\tq_j-\tq_{j+1}} 2\cosh \frac{\tp_j+\tp_{j+1}}{2} \\
    & + R^2 e^{-\tq_1-\tq_2}2\cosh \frac{\tp_1-\tp_2}{2} + R^2 e^{\tq_N+\tq_{N-1}} 2\cosh\frac{\tp_N-\tp_{N-1}}{2} + R^4  e^{-2\tq_{2}} + R^4 e^{2\tq_{N-1}}. \\
\end{split}
\end{align}

\section{Instanton partition function}\label{sec:partition function}

Let us start with a quick review on the ADHM construction of SU and SO group \cite{Nekrasov:2004vw,Haouzi:2020yxy}.

\subsection{SU group}

The ADHM construction for the $k$-instanton of $U(N)$ gauge group is given by the supersymmetric quantum mechanics of a $U(k)$ gauge theory with $N$ hypermultiplet in the fundamental representation. The ADHM data are defined by maps between the two vector spaces $\bN = \BC^{N}$ and $\bK = \BC^k$:
\begin{align}
    B_{1,2} \in \text{End}(\bK), \ I \in \text{Hom}(\bN,\bK), \ J \in \text{Hom}(\bK,\bN). 
\end{align}
along with the real and complex moment maps given by
\begin{subequations}
\begin{align}
    \mu_\BR & = \sum_{i=1}^2 \left[ B_i, B_i^\dagger \right] + II^\dagger - J^\dagger J, \\
    \mu_\BC & = [B_1,B_2] + IJ. 
\end{align}
\end{subequations}
The instanton moduli space are constructed by
\begin{align}
    \fM^{(SU)}_{N,k} = \left\{ (B_1,B_2,I,J) | \  \mu_\BR = \zeta \text{id}_{k}, \ \mu_\BC = 0 \right\}. 
\end{align}
Here $\zeta$ is the FI-parameter. The instanton partition function for $\CalN=2$ is the integration over the moduli space of instanton
\begin{align}
    \CalZ^{(SU)} = \sum_{k=0}^\infty \kq^k \CalZ_k^{(SU)}, \ \CalZ_{k}^{(SU)} = \int _ {\fM_{N,k}^{(SU)}} 1. 
\end{align}

The partition function of the supersymmetric quantum mechanics enjoys a global symmetry  
\begin{align}
    \sH = U(N) \times U(1)_\ve^2
\end{align}
of the ADHM construction of $U(N)$ gauge group subjected to $\Omega$-deformation, with its maximal torus $\sT_\sH \subset \sH$ acts naturally on the instanton moduli space $\fM^{(SU)}_{N,k}$, allowing further equivariant localization. The $\Omega$-deformation acts on the ADHM data by
\begin{align}
    (q_1,q_2) \cdot (B_1,B_2,I,J) \to (q_1B_1,q_2B_2,I,q_{1}q_2J)
\end{align}
where $q_i = e^{\ve_i}$, $i=1,2$ are the exponentiated $\Omega$-deformation parameters. 
As a result, the partition function is computed as rational function in the equivariant parameters $\xi \in \text{Lie}(\sT_\sH)$: 
\begin{align}\label{def:Z_k-int}
\begin{split}
    \CalZ_k^{(SU)} & = \int_{\fM^{(SU)}_{N,k}} ^{\sT_\sH} 1 
\end{split}
\end{align}
The integration can be evaluated by calculating residues at appropriate poles \cite{Braverman:2004vv,Nekrasov:2002qd}. Let us show some details here for better demonstration. The fix points on the moduli space is determined by the condition that the $U(k)$ and $U(N)$ transformation of the maps can be compensated by $\Omega$-deformation
\begin{align}
    g B_1 g^{-1} = q_1 B_1, \ g B_2 g^{-1} = q_2 B_2, \ g I h^{-1} = I, \ h J g^{-1} = q_{1}q_2 J
\end{align}
Here $g = e^{\hat\phi}$, $h= e^{\hat{a}}$. $\hat\phi\in \text{Lie} (\text{GL}(k))$, $\hat{a}\in \text{Lie} (\text{GL}(N))$. Denote
\[
    I = \bigoplus^N_{\alpha=1} I_\alpha, \quad J = \bigoplus^N_{\alpha=1} J_\alpha.
\]
The infinitesimal analogue of the symmetry transformation
\begin{align}\label{eq:fix-points}
\begin{split}
    & [\hat\phi,B_1] = \ve_1B_1, \ [\hat\phi,B_2] = \ve_2B_2, \ [B_1,B_2] + IJ = 0 \\
    & \hat\phi I_\alpha - I_\alpha a_\alpha = 0, \ a_\alpha J_\alpha - J_\alpha \hat{\phi} = (\ve_1+\ve_2) J_\alpha
\end{split}
\end{align}
modulo the complex gauge transformation. In the matrix form $\hat\phi = \text{diag}(\phi_1,\dots,\phi_k)$.

\paragraph{Lemma:} The instanton vector space $\bK$ satisfies the stability condition:
\begin{align}
    \bK = \BC[B_1,B_2] I(\bN)
\end{align}
when the FI-parameter $\zeta$ is positive. 
\paragraph{Proof:} Assume there exists a $\bK_\perp = \bK - \BC[B_1,B_2]I(\bN)$. We can define a projection operator $P_\perp:\bK \to \bK_\perp$ such that $P_\perp = P_\perp^2 = P_\perp^\dagger$, and $P_\perp B_{1,2} = B_{1,2} P_\perp$, $P_\perp I = 0$. Consider the projection on the real moment map
\begin{align}
    \zeta \text{id}_{\bK_\perp} = P_\perp \mu_\BR P_\perp = [(P_\perp B_1 P_\perp), (P_\perp B_1 P_\perp)^\dagger] + [(P_\perp B_2 P_\perp),(P_\perp B_2 P_\perp)^\dagger] - P_\perp J^\dagger J P_\perp
\end{align}
Taking the trace over the full vector space $\bK=\BC^k$ gives
\begin{align}
    0 \leq \zeta \Tr (\text{id}_{\bK_\perp}) = - \Tr ( (J P_\perp)^\dagger (J P_\perp) ) \leq 0 
\end{align}
This means $\bK_\perp=0$. \textbf{q.e.d}. 

\paragraph{}
Furthermore, since $\bK$ is independent of $J$ one can simply set $J=0$. The complex moment map immediately implies $[B_1,B_2]=0$. The fix points, which are the poles in \eqref{def:Z_k-int}, can be labeled by $B_1^{i-1}B_2^{j-1}I_\alpha$, $i,j=1,2,\dots$, spamming the vector space $\bK$. One can easily show that
\begin{align}
    \hat\phi B_1^{i-1}B_2^{j-1} I_\alpha = (a_\alpha + (i-1)\ve_1 + (j-1)\ve_1 ) B_1^{i-1}B_2^{j-1} I_\alpha 
\end{align}
These fix points are classified by a set of Young diagrams $\boldsymbol\lambda = (\lambda^{(1)},\dots,\lambda^{(N)})$. The partition function becomes a statistic ensemble over states labeled by $\boldsymbol\lambda$. 

The pseudo-measure associated to the instanton state $\boldsymbol\lambda$ is defined through the $\BE$-functor which converts the additive Chern character class into multiplicative class:
\begin{align}
    \BE \left[ \sum_{a} \sn_a e^{\sx_a} \right] = 
    \begin{cases}
        \displaystyle
        \prod_{a} \sx_a^{-\sn_a} & (\text{rational/4d}) \\ 
        \displaystyle
        \prod_{a} ( \sh \sx_a )^{-\sn_a} & (\text{trigonometric/5d}) \\ 
        \displaystyle
        \prod_{a} \theta(e^{-\sx_a};p) & (\text{elliptic/6d})
    \end{cases}
\end{align}
with $\sh(x) = e^{\frac{x}{2}} - e^{-\frac{x}{2}}$. In this paper we mostly only apply the trigonometric convention which corresponds to the five dimensional gauge theory. 
The pseudo measure associated to the instanton configuration $\boldsymbol\lambda=(\lambda^{(1)},\dots,\lambda^{(N)})$ is given by
\begin{align}
    \CalZ[\boldsymbol\lambda] = \BE \left[ \bN \bK^* + q_1q_{2} \bN^* \bK - P_{1}P_2 \bK \bK^* - \bM \bK^* \right].
\end{align}
$q_i=e^{\ve_i}$, $i=1,2$, are the exponentiated $\Omega$-deformation parameters with $P_i = 1-q_i$. Here we are abusing the notation of the vector spaces and their Chern characters
\begin{align}
\begin{split}
    \bN = \sum_{\alpha=1}^{N} e^{a_\alpha}, \quad \bK = \sum_{\alpha=1}^{N} \sum_{(\ri,\rj)\in \lambda^{(\alpha)}} e^{a_\alpha + (\ri-1)\ve_1 + (\rj-1)\ve_2}, \ \bM = \sum_{f=1}^{N_f} e^{m_f}. 
\end{split}
\end{align}
Given a virtual character $\bX = \sum_{a} \sn_a e^{\sx_a}$ we denote by $\bX^* = \sum_{a} \sn_a e^{-\sx_a}$ its dual virtual character. $m_f$, $f=1,\dots,N_f$, are the fundamental hypermultiplet masses.

\subsection{SO group}

The $k$-instanton moduli space of $SO(2N)$ gague group admits a standard ADHM construction. The supersymmatric quantum mechanics of the $k$-instanton moduli space is described by a $Sp(k) \subset U(2k)$ gauge theory with 1 hypermultiplet in the second rank anitsymmetric representation and $N$ hypermultiplet in the fundamental representation \cite{Nekrasov:2004vw,Benvenuti:2010pq,Marino:2004cn}. 
The ADHM data $(B_{1,2},I,J)$ for $k$-instanton moduli space are defined by
\begin{align}
\begin{split}
    & B_{1,2} = \begin{pmatrix}
        X_{1,2} & Z'_{1,2} \\ Z_{1,2} & X^T_{1,2}
    \end{pmatrix}, \\
    & J = (V,V'), \quad I^\dagger = (-V^{'*},V^*),
\end{split}
\end{align}
where $Z_{1,2}$, ${Z}'_{1,2}$ are anti-symmetric such that $B_{1,2}$ are in the anti-symmetric representation of the $Sp(k)$ satisfying \cite{itoyama1999usp}:
\begin{align}\label{eq:B-symplectic}
    B_{1,2}^T = - \BJ B_1 \BJ, \ \BJ = \begin{pmatrix}
        0 & 1_k \\ -1_k & 0
    \end{pmatrix}. 
\end{align}
In other words, $B_{1,2}\bJ$ is anti-symmetric. 
The real and complex moment maps take the form
\begin{align}\label{eq:moment-map-so}
    \mu_\BC = \begin{pmatrix}
        M_\BC & N'_{\BC} \\ N_{\BC} & -M_\BC^T 
    \end{pmatrix}, \quad
    \mu_\BR = \begin{pmatrix}
        M_\BR & N'_{\BR} \\ N_{\BR} & -M_\BR^T 
    \end{pmatrix}
\end{align}
where
\begin{subequations}
\begin{align}
    M_\BC & = [X_1,X_2] +Z'_1Z_2 - Z_2'Z_1 - V^{'T} V \\
    N_\BC & = Z_1 X_2 - X_2^T Z_1 + X_1^T Z_2 - Z_2 X_1 + V^T V  \\
    N'_\BC & = Z_1' X^T_2 - X_2 Z_1' + X_1 Z'_2 - Z_2' X_1^T - V^{'T} V'
\end{align}
\end{subequations}
and
\begin{subequations}
\begin{align}
    M_\BR & = \sum_i [X_i,X_i^T] + Z_i^* Z_i - Z_i' Z_i^{'*} + V^{'T} V^{'*} - V^{\dagger} V \\
    N_\BR & = \sum_i Z_i X_i^\dagger - X_i^* Z_i + Z_i^{'*}X_i - X_i^T Z_i^{'*} - V^T V^{'*} - V^{'\dagger} V \\
    N'_\BR & = \sum_i Z_i' X_i^* - X_i^\dagger Z_i' + Z_i^* X_i^T - X_i Z_i^* - V^{'T} V^* - V^\dagger V'
\end{align}
\end{subequations}
Notice that $N_\BC$, $N_\BC'$, $N_\BR$, and $N_\BR'$ are symmetric matrices such that $\mu_\BR \BJ$ and $\mu_\BC \BJ$ are symmetric. 
The $k$-instanton moduli space is defined by ADHM data with vanishing real and complex moment map
\begin{align}\label{def:inst-moduli-SO}
    \fM^{(SO)}_{N,k} = \left\{ (B_1,B_2,I,J) | \ \mu_{\BR} = 0, \ \mu_\BC = 0 \right\} 
\end{align}

The partition function of the supersymmetric quantum mechanics is given by an ensemble of instanton number
\begin{align}
    \CalZ^{(SO)}(\mathbf{b},\boldsymbol\ve;\kq_{SO}) = \sum_{k=0}^\infty \frac{\kq^k_{SO} }{2^k k!} \CalZ_{k}^{(SO)}
\end{align}
with the $k$-instanton pseudo-measure taking the form \cite{Nekrasov:2004vw,Haouzi:2020yxy}
\begin{align}\label{eq:SO-measure}
\begin{split}
    \CalZ_k^{(SO)}
    = & \oint \prod_{I=1}^k \frac{d\phi_I}{2\pi\ri} \frac{\sh\ve_+}{\sh\ve_1 \sh\ve_2} \frac{\sh(2\phi_I)^2\sh(2\phi_I+\ve_+)\sh(2\phi_I-\ve_+)}{\prod_{\beta=1}^N \sh \left( \frac{\ve_+}{2} \pm \phi_I \pm b_\beta \right)} \\
    & \times \prod_{I\neq J} \frac{\sh(\phi_I-\phi_J)\sh(\phi_I-\phi_J+\ve_+)}{\sh(\phi_I-\phi_J+\ve_1)\sh(\phi_I-\phi_J+\ve_2)} \times \prod_{I<J} \frac{\sh\pm(\phi_I+\phi_J)\sh(\ve_+\pm (\phi_I+\phi_J))}{\sh(\ve_1\pm (\phi_I+\phi_J))\sh(\ve_2\pm (\phi_I+\phi_J))}
\end{split}
\end{align}
Here $b_\beta$, $\beta=1,\dots,N$, are the Coulomb moduli parameters from the adjoint scalar
\begin{align}
    \Phi = \text{diag} \left\{ \begin{pmatrix}
        0 & -b_1 \\ b_1 & 0
    \end{pmatrix}, \dots, \begin{pmatrix}
        0 & -b_N \\ b_N & 0
    \end{pmatrix}
    \right\}. 
\end{align}

The contour integration can be evaluated by JK-residue technique \cite{Haouzi:2020yxy}. Remarkably in the unrefined limit $\ve_1 = -\ve_2 = \hbar$ the JK-poles can be classified by $N$ tuples of Young diagrams $\boldsymbol\lambda = (\lambda^{(1)},\dots,\lambda^{N})$ with the total number of boxes $\sum_{\beta=1}^N |\lambda^{(\beta)}| = k$ \cite{Nawata:2021dlk,Marino:2004cn,Nekrasov:2004vw}. A pole location can be specified by a box $(i,j) \in \lambda^{(\beta)}$ by 
\begin{align}
    \phi_{i,j} = b_\beta + (i-j) \hbar
\end{align}
The instanton pseudo-measure associated to the Young diagram $\boldsymbol\lambda$ reads
\begin{align}
    \CalZ^{(SO)}[\boldsymbol\lambda] = \BE \left[ 2(\bN+\bN^*) \bK + (q+q^*-2)(\bK\bK+\bK\bK^*) \right] \BI \left[ -(q^\frac{1}{2}+q^{-\frac{1}{2}}+2)\bK \right]
\end{align}
with $q=e^{\hbar}$ and the Chern characters are defined by
\begin{align}
    \bN = \sum_{\beta=1}^N e^{b_\beta}, \ \bK = \sum_{\beta=1}^N \sum_{(i,j)\in \lambda^{(\beta)}} e^{b_\beta + (i-j)\hbar }
\end{align}
Here we define another $\BI$-functor which converts the additive Chern character class into multiplicative class:
\begin{align}
    \BI \left[ \sum_a \sn_a e^{\sx_a}  \right] = \prod_a \sh(2\sx_a)^{-\sn_a}. 
\end{align}
The instanton partition function of $SO(2N)$ gauge group in the un-refined limit is an ensemble over the Young diagrams
\begin{align}\label{eq:Z-SO-unrefine}
    \CalZ^{(SO)} = \sum_{\boldsymbol\lambda} \kq_{SO}^{|\boldsymbol\lambda|} \CalZ^{(SO)} [\boldsymbol\lambda]
\end{align}

\subsection{Folding from SU to SO} 

The folding of $SU(2N)$ to $SO(2N)$ instanton can be performed on the level of moduli space. 
The vector spaces are decomposed by $\bN=\bN_+ \oplus \bN_-$ and $\bK = \bK_+ \oplus \bK_-$. 
The ADHM data $(B_1,B_2,I,J)$ for $2k$ instanton are denoted by
\begin{align}
\begin{split}
    B_{i} = \begin{pmatrix}
        B_{i,++} & B_{i,+-} \\ B_{i,-+} & B_{i,--}
    \end{pmatrix}, \quad I = \begin{pmatrix}
        I_{++} & I_{+-} \\ I_{-+} & I_{--}
    \end{pmatrix}, \quad J = \begin{pmatrix}
        J_{++} & J_{+-} \\ J_{-+} & J_{--}
    \end{pmatrix}
\end{split}
\end{align}
$B_{i,\sigma\sigma'} \in \text{Hom}(\bK_\sigma,\bK_{\sigma'})$ are $k \times k$ matrices, $I_{\sigma\sigma'} \in \text{Hom}(\bN_\sigma,\bK_{\sigma'})$ are $k \times N$ matrices, $J_{\sigma\sigma'} \in \text{Hom}(\bK_\sigma,\bN_{\sigma'})$ are $N \times k$ matrices. The real and complex moment maps are
\begin{subequations}
\begin{align}
    \mu_{\BR} & = \sum_{i=1}^2 [B_i,B_i^\dagger] + II^\dagger - J^\dagger J 
    \\
    \mu_{\BC} & = [B_1,B_2] + IJ 
\end{align}
\end{subequations}
We define moduli space which aligns with the moment maps of the $SO$ group ADHM data \eqref{eq:moment-map-so} \cite{Nekrasov:2000ih}:
\begin{align}\label{def:SU-SO-moduli}
    \fM_{2N,2k}^{(SU)} = \left\{ (B_1,B_2,I,J) | \ \mu_\BR = \zeta_{\BR} \begin{pmatrix}
        1 & 0 \\ 0 & -1
    \end{pmatrix}, \ \mu_\BC = \begin{pmatrix}
        0 & \zeta_{\BC}^+ \\ \zeta_{\BC}^- & 0
    \end{pmatrix} . \right\}
\end{align}
It's straight forward to see that the $SO$ instanton moduli space can be embedded in to such $SU$ instanton moduli space when FI-parameters $\zeta_\BR$, $\zeta_\BC^\pm$ are turned off
\begin{align}
    \fM^{(SO)}_{2N,k} \subset \fM^{(SU)}_{2N,2k}|_{\zeta_{\BR}=\zeta_\BC^\pm=0}
\end{align}
by comparing to the ADHM data defining the $k$-instanton moduli space of $SO(2N)$ group. In particular it can be obtained by restraining the $SU$ ADHM data components by
\begin{itemize}
    \item $B_{i,++} = B^T_{i,--} = X_i$. 
    \item $B_{i,+-} = Z'_{i}$ antisymmetric. 
    \item $B_{i,-+} = Z_i$ antisymmetric. 
    \item $V = \begin{pmatrix}
        J_{++} \\ J_{-+}
    \end{pmatrix} = \begin{pmatrix}
        I_{-+}^T \\ I_{--}^T
    \end{pmatrix}$. 
    \item $V' = \begin{pmatrix}
        J_{+-} \\ J_{--} 
    \end{pmatrix} = - \begin{pmatrix}
        I_{++}^T \\ I_{+-}^T
    \end{pmatrix}$.
\end{itemize}

The identification of $I$ and $J$ map indicates that the $\Omega$-deformation charge of $I$ and $J$ map needs to be identical
\begin{align}\label{eq:Omega-charge}
\begin{split}
    & (q_1,q_2) \cdot (B_{1,++},B_{1,--},B_{2,++},B_{2,--},I,J) \\
    & \to (q_1B_{1,++},q_1^{-1}B_{1,--},q_2B_{2,++},q_2^{-1}B_{2,--},q_+^{\frac{1}{2}} I, q_+^{\frac{1}{2}}J)
\end{split}
\end{align}
with the exponentiated $\Omega$-deformation parameters are
\[
    q_1=e^{\ve_1}, \ q_2=e^{\ve_2}, \ q_+ = e^{\ve_+}, \ \ve_+ = {\ve_1+\ve_2}. 
\]

The fix points on the moduli space is determined by the condition that the $Sp(k)$ transformation can be compensated by the $\Omega$-deformation as in \eqref{eq:fix-points} with $\hat\phi$ satisfying the symplectic condition: 
\begin{align}
    \hat\phi^T \BJ = -\BJ \hat\phi
\end{align}
in matrix notation $\hat\phi = (\phi_1,\dots,\phi_{2k})$. The $SO$ condition restricts $a_\beta=-a_{\beta+N}:=b_\beta$, $\beta=1,\dots,N$. The $Sp(2k)$ condition on $\hat\phi$ can be written as $(\phi_l + \phi_{l'}) \BJ_{l,l'} = 0$. The form of $\BJ$ in \eqref{eq:B-symplectic} implies the Young diagram must come in pairs $\phi_{I+k} = - \phi_I$, \ $I=1,\dots,k$.  The $SU$ group pseudo measure associated to the instanton configuration $\hat\phi$ is given by:
\begin{align}\label{eq:Z-SU-SO relation}
\begin{split}
    & \CalZ[\hat\phi] =  \BE \left[ (1+q_{+})(\bN+\bN^*)(\bK+\bK^*) - P_1P_2 (\bK+\bK^*)(\bK+\bK^*) - \bM (\bK+\bK^*) \right] 
\end{split}
\end{align}
where we denote $\bN = \sum_{\beta=1}^N e^{b_\beta}$, $\bK = \sum_{I=1}^k e^{\phi_I}$, and $\bM = \sum_{f=1}^{N_f} e^{m_f}$ labels the fundamental matters.  

We consider $B_{1,2}$ being block diagonal:
\[
    B_{i,+-} = B_{i,-+} = 0
\]
such that there are no mixture between the instantons in $\bK_+$ and $\bK_-$ spaces.
The eight sub-matrices of the real and complex moment maps become
\begin{subequations}
\begin{align}
    \mu_{\BR,++} & = \sum_{i=1}^2 [B_{i,++},B_{i,++}^\dagger] + I_{++}I_{++}^\dagger + I_{+-} I_{+-}^\dagger - I_{-+}^* I_{-+}^T - I_{--}^* I_{--}^T = \zeta_\BR {\bf 1}_k, \\
    \mu_{\BR,+-} & = I_{++}I_{-+}^\dagger + I_{+-}I_{--}^\dagger + I_{-+}^* I_{++}^T + I_{-+}^* I_{+-}^T = 0, \\
    \mu_{\BR,-+} & = I_{--}I_{+-}^\dagger + I_{-+}I_{++}^\dagger + I_{+-}^* I_{--}^* + I_{++}^* I_{-+}^T = 0, \\
    \mu_{\BR,--} & = \sum_{i=1}^2 [B_{i,--},B_{i,--}^\dagger] + I_{--}I_{--}^\dagger + I_{-+}I_{-+}^\dagger - I_{+-}^* I_{+-}^T - I_{++}^* I_{+
    +}^T = -\zeta_\BR {\bf 1}_k, 
\end{align}
\end{subequations}
and 
\begin{subequations}
\begin{align}
    \mu_{\BC,++} & = [B_{1,++},B_{2,++}] + I_{++}I^T_{-+} + I_{+-}I_{--}^T = 0, \\
    \mu_{\BC,+-} & =  -I_{++} I_{++}^T - I_{+-}I_{+-}^T = \zeta^+_\BC {\bf 1}_k, \\
    \mu_{\BC,-+} & = I_{-+}I^T_{-+} + I_{--}I_{--}^T = \zeta_\BC^- {\bf 1}_k, \\
    \mu_{\BC,--} & = [B_{1,--},B_{2,--}] -  I_{-+}I_{++}^T + I_{--}I_{+-}^T = 0.
\end{align}
\end{subequations}

\paragraph{Lemma: }The vector space $\bK = \bK_+ \oplus \bK_-$ satisfies stability condition
\begin{align}\label{eq:stability-folding}
\begin{split}
    \bK_+ & = \BC[B_{1,++}.B_{2,++}] \left[ I_{++}(\bN_+) \oplus I_{+-}(\bN_-) \right]  \\
    \bK_- & = \BC [B_{1,--}, B_{2,--}] \left[ I_{++}^*(\bN_+) \oplus I_{+-}^*(\bN_-) \right]
\end{split}
\end{align}
when $\zeta_\BR>0$. 
\paragraph{Pf: } 
It is enough to show the case that $\bK_+$ is true. 
Suppose there exists a subspace $\bK_\perp = \bK_+ - \BC[B_{1,++}.B_{2,++}] \left[ I_{++}(\bN_+) \oplus I_{+-}(\bN_-) \right]$. We can define a projection operator $P_\perp: \bK_+ \to \bK_\perp$ such that $P_\perp=P_\perp^2=P_\perp^\dagger$, and $P_\perp B_{i,++} = B_{i,++}P_\perp$, $P_\perp I_{++} = P_\perp I_{+-} = 0$. Consider the projection on the real moment map
\begin{align}
\begin{split}
    \zeta_\BR \text{id}_{\bK_\perp} 
    & = P_\perp \mu_{\BR,++} P_\perp \\
    & = \sum_{i=1}^2 [ P_\perp B_{i,++} P_\perp,(P_\perp B_{i,++} P_\perp) ^\dagger] -  P_\perp I_{-+}^* I_{-+}^T  P_\perp -  P_\perp I_{--}^* I_{--}^T  P_\perp 
\end{split}
\end{align}
Taking the trace of the full vector space $\bK_+ = \BC^k$ gives
\[
    0 \leq \zeta_{\BR} \Tr \text{id}_{\bK_\perp} = - \Tr \left( P_\perp I_{-+}^* I_{-+}^T  P_\perp +  P_\perp I_{--}^* I_{--}^T  P_\perp \right) \leq 0
\] 
This means $\bK_\perp = 0$. A similar arguemnt holds for $\bK_-$. \textbf{q.e.d.}

The stability allows us to set $I_{-+} = 0 = I_{--}$, which further implies $[B_{1,++},B_{2,++}]=0=[B_{1,--},B_{2,--}]$. 
We set $I_{+-}=0$ such that
the instanton configuration is labeled by $N$ tuples of Young diagrams $\boldsymbol\lambda=(\lambda^{(1)},\dots,\lambda^{(N)})$. The character of the vector spaces are given by
\begin{align}
\begin{split}
    \bN_\pm & = \sum_{\b=1}^N e^{\pm b_\b} \\
    \bK_+ & = \sum_{\b=1}^N \sum_{(i,j)\in \lambda^{(\b)} } e^{b_\b} q_+^{\frac{1}{2}} q_1^{i-1}q_2^{j-1} \\
    \bK_- & = \sum_{\b=1}^N \sum_{(i,j)\in \lambda^{(\b)}} e^{-b_\b} q_+^{-\frac{1}{2}} q_1^{1-i}q_2^{1-j} = \bK_+^*. 
\end{split}
\end{align}
We can denote $\bN = \bN_+ = \bN_-^*$, $\bK = \bK_+ = \bK_-^*$ 
The pseudo measure is given by
\begin{align}\label{eq:instanton-measure-folding}
\begin{split}
    \CalZ^{(SU)}[\boldsymbol\lambda]
    & = \BE\left[ (q_+^{\frac{1}{2}}+q_+^{-\frac{1}{2}}) (\bN+\bN^*)(\bK+\bK^*) - P_{12} (\bK+\bK^*) (\bK+\bK^*) - \bM (\bK+\bK^*) \right]. \\
\end{split}
\end{align}
The perturbative contribution of the partition function is given by
\begin{align}
    \CalZ^{(SU)}_{\rm pert}(\bb,\boldsymbol\ve) = \BE \left[ - \frac{(\bN+\bN^*)^2 - \bM q_+^{\frac12} (\bN+\bN^*) }{P_{12}^*} \right]
\end{align}
The combined contribution of the perturbative and instanton factor can be organized as  
\begin{align}
    \CalZ^{(SU)}[\boldsymbol\lambda] = \BE \left[ - \frac{(\bS+\bS^*)(\bS+\bS^*)^*}{P_{12}^*} + \frac{\bM q_+^{-\frac12}(\bS+\bS^*)}{P_{12}^*} \right]
\end{align}
with
\begin{align}
    \bS := \bN - P_{12} q_+^{-\frac{1}{2}}  \bK. 
\end{align}

We introduce eight fundamental hypermultiplet to the $SU$ gauge theory with masses
\begin{align}\label{def:fund-mass}
    \left\{ 0,\frac{\ve_1}{2},\frac{\ve_2}{2},\frac{\ve_+}{2}, \ri\pi, \frac{\ve_1}{2}+\ri \pi, \frac{\ve_2}{2} +\ri\pi, \frac{\ve_+}{2}+\ri\pi \right\}. 
\end{align}
The $SU(2N)$ instanton pseudo measure \eqref{eq:instanton-measure-folding} becomes double of the instanton pseudo measure of $SO(2N)$ \eqref{eq:SO-measure}: 
\begin{align}\label{eq:SU=SO^2}
    \CalZ^{(SO)}[\boldsymbol\lambda] = \BE \left[ -\frac{2\bS\bS^* + (\bS^*)^2 + \bS^2 }{2P_{12}^*} + \frac{\bM q_+^{-\frac12}(\bS+\bS^*)}{2P_{12}^*} \right], \quad 
    \CalZ^{(SU)}[\boldsymbol\lambda] = \left( \CalZ^{(SO)}[\boldsymbol\lambda] \right)^2
\end{align}
The power squares comes from the double copy of $SO(2N)$ when embedded to $U(2N)$.
The instanton partition function is an ensemble over all instanton configuration
\begin{align}
    \CalZ^{(SU)}(\bb,\boldsymbol\ve;\kq_{SU}) = \sum_{\boldsymbol\lambda} \kq_{SU}^{2|\boldsymbol\lambda|} \CalZ^{(SU)}[\boldsymbol\lambda] = \sum_{\boldsymbol\lambda} \left( \kq_{SU}^{|\boldsymbol\lambda|} \CalZ^{(SO)}[\boldsymbol\lambda] \right)^2.
\end{align}
Define modified instanton partition function:
\begin{align}\label{def:alt-SO-inst}
    \CalZ^{(SO)}(\bb;\kq=\kq_{SU}) = \sum_{\boldsymbol\lambda} \kq^{|\boldsymbol\lambda|} \CalZ^{(SO)}[\boldsymbol\lambda].
\end{align}
In the unrefined limit $\ve_1=-\ve_2=\hbar$, the character of the instanton vector space becomes
\begin{align}
    \bK = \sum_{\beta=1}^N \sum_{(i,j)\in \lambda^{(\beta)}} e^{b_\beta + (i-j)\hbar }.
\end{align}
The $SU$ instanton pseudo measure associated to the Young diagram $\boldsymbol\lambda$ becomes
\begin{align}
\begin{split}
    \CalZ^{(SU)}[{\boldsymbol\lambda}] 
    & = \left( \BE\left[ 2(\bN+\bN^*) \bK + (q+q^*-2) (\bK\bK+\bK\bK^*) \right] \BI\left[ -(q^{\frac{1}{2}}+q^{-\frac{1}{2}}+2)\bK \right] \right)^2 \\
    & = \left( \CalZ^{(SO)}[{\boldsymbol\lambda}] \right)^2
\end{split}
\end{align}
The power square comparing to the $SO(2N)$ instanton partition function pseudo-measure in the unrefined limit \eqref{eq:Z-SO-unrefine} is indicated by the double copy from the folding. This agrees with the JK-residue computation \cite{Nawata:2021dlk}.

\paragraph{}
We would like to point out here that although the unrefined limit allows the folding from $SU(2N)$ gauge theory to $SO(2N)$ on the non-perturbative level. It is not a limit one takes in the context of Bethe/Gauge correspondence. As mentioned the quantum uplift of the classical Bethe/Gauge correspondence is obtained in the NS-limit $\ve_2 \to 0$, $\ve_1 = \hbar$ rather than the unrefined limit. Nevertheless, it still provide enough information to reconstruct the type D RTL in the next section.

\section{Bethe/Gauge correspondence}\label{sec:gauge}

\subsection{$\CalN=1$ 5d gauge theory}

The correspondence between the four-dimensional $\CalN=2$ supersymmetric gauge theories and non-relativistic algebraic integrable models is extended to the quantum version in \cite{Nikita-Pestun-Shatashvili,Nikita-Shatashvili-1} by introducing the $\Omega$-deformation in the 4d gauge theories.
The low energy physics of the 4d $\CalN=2$ supersymmetric gauge theory is effectively described by a twisted super potential $\EW^{(4D)}$, an multi-valued function on the Coulomb branch defined in terms of the Nekrasov parition function in the Nekrasov-Shatashvili limit (NS-limit): 
\begin{align}
    \EW^{(4D)} (\ba,\ve_1=\hbar;\kq) = \lim_{\ve_2\to 0} \ve_2 \log \CalZ^{(4D)}(\ba,\ve_1,\ve_2;\kq).  
\end{align}
The twisted super potential of the 4d theory coincides with a 2d $\CalN=(2,2)$  supersymmetric gauged sigma model that lives on the $\BC_1$ subspace. There the two dimensional theory acquires a twisted mass $\ve_1$ from the $\Omega$-background, and the F-term equation is
then identified with the Bethe ansatz equation of the quantum integrable system. 
The stationary states of the quantum integrable system are in one-to-one correspondence to the vacuua of the effective 2d $\CalN=(2,2)$ theories. 

Such correspondence can be uplifted to in between 5d $\CalN=1$ supersymmetric gauge theories on $S^1_R \times \BR^4$ and the relativistic integrable models \cite{Nekrasov:1996cz}. 
In the NS-limit, the $\Omega$-deformed 5d theory is effectively described by a 3d $\CalN=2$ supersymmetric theory on $S^1_R\times \BC_2$, which captures the integrable nature of the system \cite{Chen:2012we,Lee:2023wbf,Kimura:2020bed,Ding:2023auy}. 

Let us start with the instanton partition function derived in the last section \eqref{def:alt-SO-inst}. 
For later convenience, 
the Coulomb moduli parameters are denoted by $\bb=\left(b_1,\dots,b_{N}\right)$. The instanton partition function is an ensemble over $N$ Young diagrams $\boldsymbol\lambda=(\lambda^{(1)},\dots,\lambda^{(N)})$:
\begin{align}\label{def:alt-SO-inst-shift}
\begin{split}
    & \CalZ^{(SO)}(\bb;\kq) = \sum_{\boldsymbol\lambda} \kq^{|\boldsymbol\lambda|} \CalZ^{(SO)}[\boldsymbol\lambda] \\
    & \CalZ^{(SO)}[\boldsymbol\lambda] = \BE \left[ - \frac{2\bS\bS^*+(\bS^*)^2+\bS^2}{2P_{12}^*} \right] \BI \left[  \frac{1}{2}(1+q_1^{\frac12}+q_2^{\frac12}+q_+^{\frac12})\left( \frac{q_+^{\frac{1}{2}} \bS}{P_{12}} + \frac{q_+^{-\frac{1}{2}}\bS^*}{P_{12}^*} \right) \right].
\end{split}
\end{align}
Denote
\begin{align}\label{def:x_b}
    x_{\b i} = b_\b + \left(i-\frac12\right)\ve_1 +  \lambda^{(\b)}_i \ve_2, \ \bX = \sum_{(\b i)} e^{x_{\b i}}, \ \bS = P_1 q_1^{-\frac12} \bX
\end{align}
for a given instanton configuration $\boldsymbol\lambda$. The $SO$ instanton pseudo-measure can be expressed as 
\begin{align}
\begin{split}
    \CalZ_{SO}[\boldsymbol\lambda]
    = & \ \BE \left[ \frac{-2P_1\bX\bX^*+P_1(\bX^*)^2+P_1\bX^2}{2P_2^*} \right] \BI \left[ \frac{1}{2}(1+q_1^{\frac12}+q_2^{\frac12}+q_+^{\frac12}) \left( \frac{q_2^{\frac{1}{2}}\bX}{P_{2}} + \frac{q_2^{-\frac12}\bX^*}{P_2^*} \right) \right] \\
    = & \prod_{(\b i)\neq (\b'i')} \frac{\Gamma_{q_2}\left( \frac{x_{\b i}-x_{\b' i'}-\ve_1}{\ve_2} \right) }{\Gamma_{q_2}\left( \frac{x_{\b i}-x_{\b' i'}}{\ve_2} \right)}  \prod_{(\b i),(\b'i')} \left[ \frac{\Gamma_{q_2}\left( \frac{x_{\b i}+x_{\b' i'}}{\ve_2} \right) }{\Gamma_{q_2}\left( \frac{x_{\b i}+x_{\b' i'}-\ve_1}{\ve_2} \right)} \frac{\Gamma_{q_2}\left( \frac{-x_{\b i}-x_{\b' i'}}{\ve_2} \right) }{\Gamma_{q_2}\left( \frac{-x_{\b i}-x_{\b' i'}-\ve_1}{\ve_2} \right)}     \right]^{\frac12} \\
    & \times \prod_{(\b i)}\left[  \frac{ \G_{q_2^{2}} \left( \frac{x_{\b i}+\frac{\ve_2}{2}}{\ve_2} \right) \G_{q_2^{2}} \left( \frac{x_{\b i}+\frac{\ve_2}{2}-\frac{\ve_1}{2}}{\ve_2} \right)  \G_{q_2^{2}} \left( \frac{x_{\b i}}{\ve_2} \right) \G_{q_2^{2}} \left( \frac{x_{\b i}-\frac{\ve_1}{2}}{\ve_2} \right) }{ \G_{q_2^{2}} \left( \frac{-x_{\b i}+\frac{\ve_2}{2}}{\ve_2} \right) \G_{q_2^{2}} \left( \frac{-x_{\b i}+\frac{\ve_2}{2}-\frac{\ve_1}{2}}{\ve_2} \right)  \G_{q_2^{2}} \left( \frac{-x_{\b i}}{\ve_2} \right) \G_{q_2^{2}} \left( \frac{-x_{\b i}-\frac{\ve_1}{2}}{\ve_2} \right) } \right]^{\frac{1}{2}}
\end{split}
\end{align}
Here the $\G_q$-function is the $q$-gamma function defined by
\begin{align}
\begin{split}
    & \Gamma_q(x) =  q^{-\frac{(x-1)x}{4}} \prod_{n=0}^\infty \frac{1-q^{n+1}}{1-q^{n+x}} = q^{-\frac{(x-1)x}{4}} \frac{(q;q)_\infty}{(q^x;q)_\infty}, 
\end{split}
\end{align}
\[
    \frac{\G_q(x+1)}{\G_q(x)} = q^{\frac{x}{2}} - q^{-\frac{x}{2}}.
\]

Now we consider the Nekrasov-Shatashvili limit $\ve_2 \to 0$ where the $\CalN=2$ super-Poincar\'{e} symmetry is restored and the five dimensional $\CalN=1$ theory is now effectively described by the three dimensional $\CalN=2$ theory. Precisely speaking, the twisted superpotential of the two theories coincide when evaluated on their corresponding vacua \cite{Chen:2012we,HYC:2011,Dorey:2011pa}. 
In this limit, we can approximate the $q$-gamma function using the Stirling formula for $q$-gamma function \cite{moak1984q}: 
\begin{align}
\begin{split}
    \lim_{\log q_2\to 0} \log \G_{q_2^l} \left( \frac{x}{\log q_2} \right)
    & = \sum_{n=0}^\infty \log (1-q_2^{l(n+1)}) - \frac{x^2}{4\ve_2} l + \frac{1}{l\ve_2} \text{Li}_2(e^{lx}) + \CalO(\ve_2). 
\end{split}
\end{align}
$\text{Li}_2$ is dilogarithm. 
The instanton partition function has the asymptotics
\begin{align}\label{def:twist-super-potential}
    \lim_{\ve_2 \to 0} \CalZ^{(SO)}(\bb,\ve_1,\ve_2;\kq) = e^{-\frac{\EW(\bb,\hbar=\ve_1;\kq)}{\ve_2}} + \CalO(\ve_2)
\end{align}
The twisted superpotential $\EW(\ba,m_f,\ve_1;\kq)$ is specified to
\begin{align}
\begin{split}
    \EW(\bb,\hbar;\kq) 
    = & \ \frac{1}{2} \sum_{(\b i)\neq (\b'i')} \text{Li}_2 \left( e^{\pm x_{\b i} \pm x_{\b' i'}-\hbar} \right) - \frac{\left( \pm x_{\b i} \pm x_{\b' i'}-\hbar \right)^2}{4} \\
    & \qquad \qquad \quad - \text{Li}_2 \left( e^{\pm x_{\b i} \pm x_{\b' i'}} \right) + \frac{\left( \pm x_{\b i} \pm x_{\b' i'} \right)^2}{4} \\
    & + \sum_{(\b i)} x_{\b i} \log \kq + \text{Li}_2 ( 2x_{\b i}) - \text{Li}_2 (-2x_{\b i} - \hbar) 
\end{split}
\end{align}

The twisted superpotential $\EW(\bb,\hbar;\kq)$ recovers the effective potential of pure 3-dimensional $\CalN=2$ theory \cite{Kimura:2020bed,Ding:2023auy,Aganagic:2013tta,Yoshida:2014ssa}. 
The ensemble over instanton configuration $\CalZ^{(SO)}[\boldsymbol\lambda]$ is dominated by the limit shape instanton configuration $\boldsymbol\Lambda=(\Lambda^{(1)},\dots,\Lambda^{(N)})$ defined by
\begin{align}\label{def:limit-shape}
    \kq^{|\boldsymbol\Lambda|} \CalZ^{(SO)}[\boldsymbol\Lambda] = e^{-\frac{\EW(\bb,\hbar;\kq)}{\ve_2}}. 
\end{align}
The vacuum equation of the twisted superpotential is given by
\begin{align}
    \exp \left( \frac{\p}{\p x_{\b i}} \EW(\bb,\hbar;\kq) \right) = 1
\end{align}
which reads
\begin{align}\label{eq:BAE}
    \kq {\sh (2x_{\b i})^2 \sh (2x_{\b i}-\hbar)^2 } \prod_{(\b'i')\neq (\b i)} \frac{\sh(x_{\b i} \pm x_{\b'i'}-\hbar)}{\sh(x_{\b i} \pm x_{\b'i'}+\hbar)}  = 1. 
\end{align}
We recovered the Bethe ansatz equation of type D relativistic Toda lattice \cite{sklyanin1988boundary,kuznetsov1997separation,kuznetsov1996quantum}. 

\subsection{\textit{qq}-character}

The introduction of defects in the gauge theory turns out to be an useful tool studying the quantum version of the Bethe/Gauge correspondence \cite{Nikita:V,Lee:2020hfu,Jeong:2021rll,Jeong:2023qdr,Jeong:2024hwf}.
It is known that the Nekrasov-Shatashvili free energy, which works excellently for 4d, is not enough to establish 5d Bethe/Gauge correspondence. The correct quantization requires including a tower of non-perturbative effects \cite{Grassi:2014zfa,Franco:2015rnr,Hatsuda:2015qzx,Grassi:2017qee}.
Our approach here relies on the co-dimensional four observable known as the \emph{qq}-character \cite{Nikita:I} -- a quantum-quantum uplift of the Seiberg-Witten curve. Here the fundamental $qq$-character is a line defect warpping on the compactified $S^1_R$ in the 5d gauge theory \cite{Kim:2016qqs}.  

The fundamental $qq$-character observable can be obtained by considering a variation on the bulk instanton configuration \eqref{def:alt-SO-inst}. This is done by adding an extra instanton at $X=e^x$, $\bK \to \bK + X$, $\bS \to \bS - P_{12}q_+^{-\frac{1}{2}}X$. The instanton pseudo measure varies by
\begin{align}
\begin{split}
    & \kq \frac{\CalZ^{(SO)}\left[\bS-P_{12}q_+^{-\frac12}X\right]}{\CalZ^{(SO)}[\bS]} \\
    & = \BE \left[ - \frac{2(\bS-P_{12}q_+^{-\frac{1}{2}}X)(\bS^*-P_{12}^*q_+^{\frac{1}{2}}X^*)+(\bS-P_{12}q_+^{-\frac{1}{2}}X)^2+(\bS^*-P_{12}^*q_+^{\frac{1}{2}}X^*)^2}{2P_{12}^*} + \frac{2\bS\bS^*+\bS^2+(\bS^*)^2}{2P_{12}^*}  \right] \\
    & \quad \times \BI \left[ -\frac12 (1+q_1^{\frac12}+q_2^{\frac12}+q_+^{\frac12}) (\bK+\bK^*+ X+X^*) + \frac12 (1+q_1^{\frac12}+q_2^{\frac12}+q_+^{\frac12}) (\bK+\bK^*) \right] \\
    & = \BE \left[  q_+^{\frac{1}{2}} X (\bS- P_{12} q_+^{-\frac{1}{2}}X)^* + \bS q_+^{\frac{1}{2}} X^* + q_+^{\frac{1}{2}} X\bS + q_+^{\frac{1}{2}}X^*\bS^* - P_{12} \frac{(X^*)^{2}}{2} - P_{12} \frac{X^2}{2} \right] \\
    & \quad \times \BE\left[ - \frac12 (1+q_1+q_2+q_+)(X^2 + (X^*)^2) \right] \\
    & = \frac{\kq P(x)}{Y\left(x+\frac{\ve_+}{2}\right)Y\left(-x+\frac{\ve_+}{2}\right)Y\left(x-\frac{\ve_+}{2}\right)Y\left(-x-\frac{\ve_+}{2}\right)}.
\end{split}
\end{align}
Here we define the $Y$-function by 
\begin{align}
\begin{split}
    Y(x)[\boldsymbol\lambda] 
    & = \BE \left[ -X \bS^* \right] = \BE \left[ - e^x \left( \bN-q_+^{-\frac{1}{2}}P_{12}\bK \right)^* \right] \\
    & = \prod_{\beta=1}^N \sh \left( x - b_\b \right) \prod_{(i,j)\in \lambda^{(\b)}} \frac{\sh(x-b_\b-i\ve_1-(j-1)\ve_2) \sh(x-b_\b-(i-1)\ve_1-j\ve_2)}{\sh(x-b_\b-(i-1)\ve_1-(j-1)\ve_2 ) \sh(x-b_\b-i\ve_1-j\ve_2 )} \\
    & = \prod_{\b=1}^N \sh \left(x-b_\b-\ve_1\lambda_1^{(\b),T} \right) \prod_{i=1}^{\lambda^{(\b)}} \frac{\sh\left( x-b_\b - (i-1)\ve_1 - \ve_2 \lambda^{(\b)}_i \right)}{\sh\left( x-b_\b - i\ve_1 - \ve_2 \lambda^{(\b)}_i \right)}, 
\end{split}
\end{align}
and 
\begin{align}
\begin{split}
    Y(-x) & = \BE \left[ -X^* \bS^* \right] \\
    & = \prod_{\beta=1}^N \sh \left( -x - b_\b \right) \prod_{(i,j)\in \lambda^{(\b)}} \frac{\sh(-x-b_\b-i\ve_1-(j-1)\ve_2) \sh(-x-b_\b-(i-1)\ve_1-j\ve_2)}{\sh(-x-b_\b-(i-1)\ve_1-(j-1)\ve_2 ) \sh(-x-b_\b-i\ve_1-j\ve_2 )} \\
    & = (-1)^N \prod_{\b=1}^N \sh \left(x+b_\b+\ve_1\lambda_1^{(\b),T} \right) \prod_{i=1}^{\lambda^{(\b)}} \frac{\sh\left( x+b_\b + (i-1)\ve_1 + \ve_2 \lambda^{(\b)}_i \right)}{\sh\left( x+b_\b + i\ve_1 + \ve_2 \lambda^{(\b)}_i \right)}. \\
\end{split}
\end{align}
We also have
\[
    P(x) = \sh(2x)^2 \sh(2x-\ve_+)\sh(2x+\ve_+).
\]
We construct the fundamental $qq$-character as  
\begin{align}\label{def:qq-character}
    \EX(x)[\boldsymbol\lambda] = Y\left(x+\frac{\ve_+}{2}\right)[\boldsymbol\lambda] Y\left(-x-\frac{\ve_+}{2}\right)[\boldsymbol\lambda] + \frac{\kq P(x)}{Y\left(x-\frac{\ve_+}{2}\right)[\boldsymbol\lambda]Y\left(-x+\frac{\ve_+}{2}\right)[\boldsymbol\lambda]}
\end{align}
whose expectation value 
\begin{align}
\begin{split}
    T(x) & = \langle \EX(x) \rangle = \frac{1}{\CalZ^{(SO)}} \sum_{\boldsymbol\lambda} \kq^{|\boldsymbol\lambda|} \CalZ^{(SO)}[\boldsymbol\lambda] \EX(x)[\boldsymbol\lambda] \\
    & = E_0 X^N + E_1 X^{N-1} + \cdots + E_{2N}X^{-N}, 
    E_0 = 
    \begin{cases}
        1+\kq_{SU}, & N=4; \\
        1, & N>4.
    \end{cases}
\end{split}
\end{align}
is an analytic function in $x$ as the poles of the $qq$-character are canceled in the ensemble \cite{Nikita:II}. 
In short words, the poles of $x$ coming from the individual $qq$-character function $\EX(x)[\boldsymbol\lambda]$ are canceled in the ensemble. 
The Seiberg-Witten curve of the supersymmetric gauge theory can be recovered by turning off the $\Omega$-deformation $\ve_{1,2} \to 0$
\begin{align}
    T(x) = Y(x)Y(-x) + \frac{\kq P(x)}{Y(x)Y(-x)}. 
\end{align}

\begin{figure}[h]
    \centering
    \begin{tikzpicture}[scale=0.6, every node/.style={scale=1.2}]
    \begin{scope}
    \draw[thick] (1,1) -- (-1,1);
    \draw[thick] (1,-1) -- (-1,-1);
    \draw[thick] (2,2) -- (-2,2);
    \draw[thick] (2,-2) -- (-2,-2);
    \draw[thick] (3,2.5) -- (2,2) -- (1,1) -- (1,-1) -- (2,-2) -- (3,-2.5);
    \draw[thick] (-3,2.5) -- (-2,2) -- (-1,1) -- (-1,-1) -- (-2,-2) -- (-3,-2.5);
    \draw[blue,thick] (-4.5,0) -- (4.5,0);

    \node[above] at (0,2) {$\vdots$};
    \node[below] at (0,-2) {$\vdots$}; 
    \node[right] at (3.7,2) {5-brane web};
    \node[above] at (-3,0) {D5};

     \draw[->,thick] (5,-1) -- (5,-0);
    \draw[->,thick] (5,-1) -- (6,-1);
    \draw[->,thick] (5,-1) -- (4.3,-1.7);

    \node[above] at (5,0) {$\bx_9$};
    \node[right] at (6,-1) {$\bx_6$};
    \node[left] at (4.3,-1.7) {$\bx_5$};
    
    \node[above] at (0,-5) {(a)};
    
    \end{scope}
    \begin{scope}[xshift = 12cm]
    \draw[thick] (1,1) -- (-1,1);
    \draw[thick] (1,-1) -- (-1,-1);
    \draw[thick] (2,2) -- (-2,2);
    \draw[thick] (2,-2) -- (-2,-2);
    \draw[thick] (3,2.5) -- (2,2) -- (1,1) -- (1,-1) -- (2,-2) -- (3,-2.5);
    \draw[thick] (-3,2.5) -- (-2,2) -- (-1,1) -- (-1,-1) -- (-2,-2) -- (-3,-2.5);
    \draw[red,thick] (-1,0) -- (-4,-3);
    \draw[blue,thick] (-4.5,-3) -- (4.5,-3);

    \node[above] at (0,2) {$\vdots$};
    \node[left] at (-2,-0.7) {D3};
    \node[below] at (0,-2) {$\vdots$}; 
    \node[below] at (3,-3) {D5};

    \draw[->,thick] (5,-1) -- (5,-0);
    \draw[->,thick] (5,-1) -- (6,-1);
    \draw[->,thick] (5,-1) -- (4.3,-1.7);

    \node[above] at (5,0) {$\bx_9$};
    \node[right] at (6,-1) {$\bx_6$};
    \node[left] at (4.3,-1.7) {$\bx_5$};

    \node[above] at (0,-5) {(b)};
    \end{scope}
    \end{tikzpicture}
    \caption{D3-brane creation and higgsing construction of $Q$-observable. (a) One of the D5-branes is aligned across the two NS5-branes, enabling the D5-brane to move along the $x^5$-direction. (b) A D3-brane ending on one of the NS5-branes is created as a result of the transition.}
    \label{fig:Q-operator}
\end{figure}
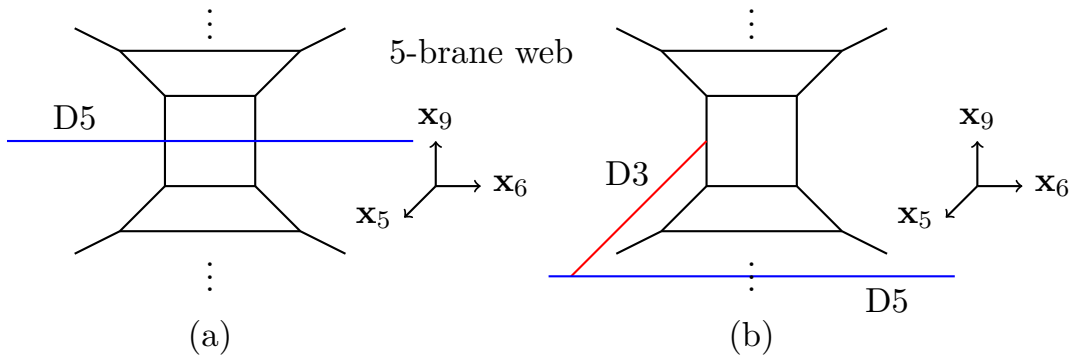

The Bethe ansatz equation that we obtained from the twisted super potential can also be obtained from the $qq$-character. 
In the NS-limit $\ve_2 \to 0$, $\ve_1 = \hbar$ where the instaotn partition function is dominated by limit shape configuration \eqref{def:limit-shape}. we define the $Q$-function 
\begin{subequations}
\begin{align}\label{def:Q-obs}
    & Y(x)[\boldsymbol\Lambda] = \frac{Q\left(x+\frac{\ve_1}{2}\right)}{Q\left(x-\frac{\ve_1}{2}\right)}, \ Q(x) = \prod_{\b=1}^{N} \prod_{i=1}^\infty \sh(x-x_{\b i}), \\
    \implies & Y(-x)[\boldsymbol\Lambda] = (-1)^N \frac{Q\left(-x+\frac{\ve_1}{2}\right)}{Q\left(-x-\frac{\ve_1}{2}\right)}, 
\end{align}
\end{subequations}
Here $x_{\b i}$ is defined in \eqref{def:x_b} with the instanton configuration as limit shape configuration $\boldsymbol\Lambda$. 
The physical meaning of $Q(x)$ is a canonical surface defect wrapping $\BR^2_{12}=\BC_1 \subset S^1 \times \BR^4$ in the 5d supersymmetric gauge theory. See Fig.~\ref{fig:Q-operator} for an illustration. 
A three dimensional $\CalN=2$ theory lives on the canonical surface defect whose twisted superpotential coincides with the 5d $\CalN=1$ superpotential \cite{Kimura:2020bed,Chen:2012we}.
The expectation value of $qq$-character \eqref{def:qq-character} in the NS-limit becomes
\begin{align}
\begin{split}
    T\left(x  \right) 
    & = \left\langle \EX\left(x\right) \right\rangle = \EX\left(x\right)[\boldsymbol\Lambda] \\
    & = Y\left(x+\frac{\hbar}{2}\right)[\boldsymbol\Lambda] Y\left(-x-\frac{\hbar}{2}\right)[\boldsymbol\Lambda] + \frac{\kq P\left(x \right) }{Y\left(x-\frac{\hbar}{2}\right)[\boldsymbol\Lambda]Y\left(-x+\frac{\hbar}{2}\right)[\boldsymbol\Lambda]} \\
    & = \frac{Q(x+\hbar)Q(-x-\hbar)}{Q(x) Q(-x)} + \kq P\left(x \right) \frac{Q(x-\hbar)Q(-x+\hbar)}{Q(x)Q(-x)}
\end{split}
\end{align}
We find the $Q$-function satisfying the Baxter T-Q equation by multiplying both side with $Q(x)Q(-x)$:
\begin{align}
    T(x)Q(x)Q(-x) = Q(x+\hbar)Q(-x-\hbar) + \kq P\left(x \right) Q(x-\hbar)Q(-x+\hbar). 
\end{align}
At the zeros of $Q(x)$ or $Q(-x)$, the vanishing right hand side requires  
\begin{align}
\begin{split}
    & -1 = \kq P\left( \pm x_{\b i}  \right) \frac{Q(x_{\b i}-\hbar)Q(-x_{\b i}+\hbar)}{Q(x_{\b i}+\hbar)Q(-x_{\b i}-\hbar)} \\
    & \implies \kq \sh(2x_{\b i})^2 \sh(2x_{\b i}-\hbar)^2 \prod_{(\beta' i') \neq (\b i)} \frac{\sh(x_{\b i} \pm x_{\beta' i'}-\hbar)}{\sh(x_{\b i}\pm x_{\beta'i'}+\hbar)} = 1
\end{split}
\end{align}
This is the same Bethe ansatz equation that we obtained from the twisted superpotential \eqref{eq:BAE}.

In the context of integrable system through Bethe/Gauge correspondence, $Q$-observable is the eigen function of Baxter $Q$-operator. Its zeros are the Bethe roots of the integrable system. See \cite{Jeong:2024hwf,Jeong:2024mxr} for the rules of $Q$-observables in both the gauge theory and integrable systems.

\subsection{Monodromy defect}

The four dimensional theory with co-dimensional two monodromy surface defect can be viewed as a theory on an orbifold. The localization technique extended to the computation of defect instanton partition function and also the expectation value of some observables. We in particular focus on the class of $qq$-observables \cite{Nikita:I}, which are fractionalized in the presence of surface defect. The main statement in \cite{Nikita:II} states certain vanishing condition for the $qq$-character. These vanishing conditions, called \emph{non-perturbative Dyson-Schwinger equation}, are useful for deriving the KZ-equations satisfied by the defect partition function. In addition, in the NS-limit $\ve_2 \to 0$, the Dyson-Schwinger equation becomes the Schr\"{o}dinger-type equation of the integrable system satisfied by the defect partition function.

\paragraph{}
The co-dimensional two defect is introduced in the form of an $\BZ_L$-orbifold acting on $\BR^4 = \BC_1 \times \BC_2$ by $(\bz_1,\bz_2) \to (\bz_1,\zeta\bz_2)$ with $\zeta^L = 1$. The orbifold modifies the ADHM construction and creates a chain-saw quiver structure \cite{Kanno:2011fw}. 
Such defect is characterized by a coloring function $c:[N] \to \BZ_L$ that assigns the representation $\CalR_{c(\b)}$ of $\BZ_L$ to each color $\b=1,\dots,N$. 
We use $\CalR_\o$ to denote the one dimensional complex irreducible representation of $\BZ_L$, where the generator $\zeta$ is represented by the multiplication of $\exp\left( \frac{2\pi \ri \o}{L} \right)$ for $\o\equiv\o+L$. In general one can consider any $\BZ_L$. A surface defect is called \emph{regular/full-type surface defect} when $L=2N$ and the coloring function $c(\alpha)$ is bijective. 

Hereafter, we will consider the case $L=2N$ and coloring function of the form
\begin{align}\label{def:color_function}
    c(\b) = \b-1. 
\end{align}
The reader might notice that there are only $N$ Coulomb moduli parameters $\bb=(b_1,\dots,b_N)$ and conclude that the monodromy defect considered here is not regular. 
Remember that $SO(2N)$ can be embedded in $U(2N)$ by restricting $N$ of the $2N$ Coulomb moduli parameters $\ba=(a_1,\dots,a_{2N})$ 
\[
    a_{\b} = - a_{2N-\b+1} = b_\b, \quad \b=1,\dots,N. 
\]


On the level of the $U(2N)$ gauge theory, the coloring function $c:[2N]\to \BZ_{2N}$ is bijective. Hence we conclude the monodromy defect is regular. 



\begin{figure}
    \centering
    \includegraphics[width=0.6\textwidth]{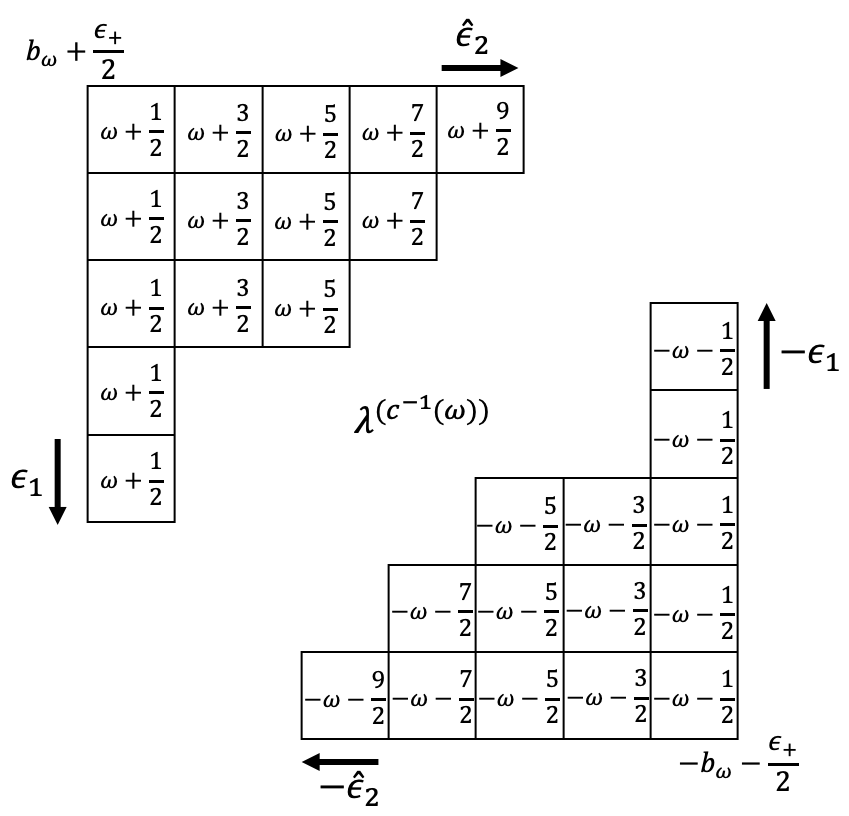}
    \caption{A Young diagram $\lambda^{(\b)}$ is two-color in the presence of orbifold.}
    \label{fig:K-mirror}
\end{figure}

The folding from $SU$ to $SO$ results in several changes on how the instanton Young diagrams are colored. First, since a Young diagram $\lambda^{(\b)}$ labels instanton configuration generated by
\[
    \bK_+ = \BC[B_{1,++},B_{2,++}] I_{++}(\bN_+)
\]
in \eqref{eq:stability-folding}. The $\Omega$-deformation charge of the ADHM data $I$ is $(q_1,q_2) \cdot I \to q_+^{\frac12}I$ \eqref{eq:Omega-charge}. 
This leads to the fact that the boxes in a colored Young diagram is in the $\CalR_{\o+\frac12}$ representation, with the additional $\frac12$ shift comes from $I$. 
The complex exponentiated gauge coupling $\kq$ fractionalized to $2N$ fractional coupling $(\kq_{\o+\frac12})_{\o=0}^{2N-1}$: 
\begin{align}
    \kq = \prod_{\o=0}^{2N-1} \kq_{\o+\frac12}, \ \kq_{\o+\frac12+2N} = \kq_{\o+\frac12}. 
\end{align}
The fractional coupling $\kq_{\o+\frac12}$ is assigned to the $\CalR_{\o+\frac12}$ representation of the $\BZ_{2N}$-orbifold. 
The boxes in the Young diagrams are colored by the representation $\CalR_\o$ based on their position. 
\begin{align}\label{def:k_w}
\begin{split}
    & \EK_{\o+\frac12}^+ = \left \{ (\b,(i,j)) \mid \b \in [N], \ (i,j)\in \lambda^{(\b)}, \ c(\b)+j-\frac12 = \omega \ \text{mod } 2N \right\}, \\
    & \EK_{\o+\frac12}^- = \left \{ (\b,(i,j)) \mid \b \in [N], \ (i,j)\in \lambda^{(\b)}, \ -c(\b)-j+\frac12 = \omega \ \text{mod } 2N \right\}, \\
    & k_{\o+\frac12}^+ = \left|\EK_{\o+\frac12}^+\right| , \quad k_{\o+\frac12}^- = \left|\EK_{\o+\frac12}^-\right|
\end{split}
\end{align}
denote the number of squares in a colored Young diagram that is in the $\CalR_{\omega+\frac12}$ representation of $\BZ_{2N}$ orbifold. Note that in the context of $U(2N)$--where we embed $SO(2N)$-- a single Young diagram $\lambda^{(\b)}$ is used for both $a_\b=b_\b$ and $a_{2N-\b+1}=-b_\b$. A square in the Young diagram is two-color. See Fig.~\ref{fig:K-mirror} for illustration.
The symmetry leads to
\begin{align}\label{eq:K_w symmetry}
    \EK_{\o+\frac12}^+ = \EK_{2N-\o-\frac12}^-.
\end{align} 
for all $\o=0,\dots,2N-1$. 

Here raise the question: how do we count the fractional instanton counting based on a given colored Young diagram $\hat{\boldsymbol\lambda}$? There are two potential options: 
\begin{itemize}
    \item Option 1: Count $\EK^+_{\o+\frac12}$ (or $\EK^-_{-\o-\frac12}$) only. 
    \item Option 2: Count both $\EK^\pm_{\o+\frac12}$, with each colored boxes count half instanton in $\EK^+_{\o+\frac12}$, and half in $\EK^-_{2N-\o-\frac12}$.  
\end{itemize}
As \eqref{eq:K_w symmetry} will cause a mixture between $\kq_{\o+\frac12}$ and $\kq_{2N-\o-\frac12}$ in later computation. Here we choose the first option.

The defect partition function is the integration over the $\BZ_{2N}$ invariant fields
\begin{align}\label{eq:orbi-fold-inst}
\begin{split}
    & \hat\CalZ^{(SO)} = \sum_{\hat{\boldsymbol\lambda}} \prod_{\o=0}^{2N-1} \kq_{\o+\frac12}^{k^+_{\o+\frac12}} \hat\CalZ^{(SO)}[\hat{\boldsymbol\lambda}] \\
    & \hat\CalZ^{(SO)}[\hat{\boldsymbol\lambda}] 
    = \BE \left[ -\frac{(\hat{\bS}_++\hat{\bS}_-)(\hat{\bS}_++\hat{\bS}_-)}{2\hat{P}_{12}^*}  \right]^{\BZ_{2N}}  \BI \left[  \frac{1}{2}(1+q_1+\hat{q}_2+\hat{q}_+)\left( \frac{\hat{q}_+^{\frac{1}{2}}\hat\bS}{\hat{P}_{12}} + \frac{\hat{q}_+^{-\frac{1}{2}}\hat\bS^*}{\hat{P}_{12}^*} \right) \right]^{\BZ_{2N}} \\
\end{split}
\end{align}
Here $\hat\bS_+ = \hat\bS = \hat\bS_-^*$. This separation will turn out to be very useful.

The $\Omega$-deformation parameters are charged with the orbifold action
\begin{align}
    \hat{q}_1 = q_1 \CalR_0, \ \hat{q}_2 = q_2^{\frac{1}{2N}} \CalR_1, \ \hat{P}_1 = (1-q_1) \CalR_0, \ \hat{P}_2 = \CalR_0 - q_2^{\frac{1}{2N}} \CalR_1.
\end{align}
All the ADHM data can be written in terms of the shifted parameters
\[
    \hat\bN_+ = \sum_{\o=0}^{N-1} \bN_{+,\o+\frac12} q_2^{\frac{\o+\frac12}{2N}} \CalR_{\o}, \quad \bN_{+,\o+\frac12} = \sum_{\b\in c^{-1}\left(\o\right)} e^{\tilde{b}_{\o}}, 
\]
\[
    \hat\bK_+ = \sum_{\o=0}^{2N-1} \bK_{+,\o+\frac12} q_2^{\frac{\o+\frac12}{2N}} \CalR_{\o+\frac12}, \quad \bK_{+,\o+\frac{1}{2}} = \sum_{\b=1}^N  \sum_{J=0}^\infty \sum_{\underset{c(\b)+j-\frac12=\omega+\frac{1}{2}+NJ}{{(i,j)\in \lambda^{(\alpha)}}}} e^{\tilde{b}_{c(\b)}} q_1^{i-\frac12}q_2^J
\]
\[
    \hat\bS_+ = \sum_{\o=0}^{2N-1} \bS_{+,\o} q_2^{\frac{\o}{2N}} \CalR_{\o}, \quad \bS_{+,\o} = \bN_{+,\o+\frac12} - P_1 \bK_{+,\o+\frac12} + q_2^{\d_{\o,0}} P_1 \bK_{+,\o-\frac{1}{2}}
\]
\[
    \hat\bN_- = \hat\bN_+^*, \quad \hat\bK_- = \hat\bK_+^*, \ \quad \hat\bS_- = \hat\bS_+^*. 
\]
Here we use the short handed notation $\bN_{\o}=0$ for $\o>N-1$. 
The bulk instanton configuration is identified with $\o=2N-1$ colored partition
\begin{align}
    \sum_{\o=0}^{2N-1} \bS_{+,\o} = \bN - P_{12} \bK_{+,2N-\frac12} = \bS
\end{align}
We define a projection $\rho(\hat{\boldsymbol\lambda}) = \boldsymbol\lambda$ from the colored partition $\hat{\boldsymbol\lambda}$ to the bulk partition $\boldsymbol\lambda$ \cite{Lee:2020hfu,Jeong:2024mxr},
\[
    \lambda^{(\b)}_i = \left\lfloor \frac{\hat\lambda^{(\b)}_i+c(\b)}{2N} \right\rfloor, \ \a=1,\dots,N; \ i=1,2,\dots
\]
The colored partition function can be reorganized into
\begin{align}
    \hat\CalZ^{(SO)} = \sum_{\boldsymbol\lambda} \kq^{|\boldsymbol\lambda|} \CalZ^{(SO)}[\boldsymbol\lambda] \sum_{\hat{\boldsymbol\lambda}\in \rho^{-1}(\boldsymbol\lambda)} \prod_{\o=0}^{2N-1} u_{\o+\frac12}^{k^+_{\o-\frac12} - k^+_{\o+\frac12} } \hat\CalZ^{(SO)}_{\rm surface}[\hat{\boldsymbol\lambda}] = \langle \psi \rangle \CalZ^{(SO)}
\end{align}
The monodromy defect observable $\psi[\boldsymbol\lambda]$ is defined by
\begin{align}\label{def:psi}
    \psi[\boldsymbol\lambda] = \sum_{\hat{\boldsymbol\lambda}\in \rho^{-1}(\boldsymbol\lambda)} \prod_{\o=0}^{2N-1} u_{\o+\frac12}^{k^+_{\o-\frac12} - k^+_{\o+\frac12} } \hat\CalZ^{(SO)}_{\rm surface}[\hat{\boldsymbol\lambda}]. 
\end{align}


\subsection{Fractional \textit{qq}-character}

We consider the variation of the defect instanton pseudo-measure by adding an instanton based on an given instanton configuration $\hat{\boldsymbol\lambda}$ at $\hat{X}=e^xq_2^{\frac{\o+\frac{1}{2}}{2N}}\CalR_{\o+\frac12}$. 
The defect pseudo measure \eqref{eq:orbi-fold-inst} is changed by
\begin{align}
\begin{split}
    & \kq_{\o+\frac{1}{2}} \frac{\hat\CalZ^{(SO)}[\hat\bS-\hat{P}_{12}\hat{q}_+^{-\frac{1}{2}}\hat{X}]}{\hat\CalZ^{(SO)}[\hat\bS]} \\
    & = \kq_{\o+\frac{1}{2}} \BE \left[  \hat{q}_+^{\frac{1}{2}} \hat{X} (\hat\bS- \hat{P}_{12} \hat{q}_+^{-\frac{1}{2}}\hat{X})^* + \hat\bS \hat{q}_+^{\frac{1}{2}} \hat{X}^* + \hat{q}_+^{\frac{1}{2}} \hat{X}\hat{\bS} + \hat{q}_+^{\frac{1}{2}}\hat{X}^*\hat\bS^* - (1+\hat{q}_+) (\hat{X}^2+(\hat{X}^*)^2) \right]^{\BZ_{2N}} \\
    & = \frac{\kq_{\o+\frac{1}{2}}  P_\o(x)}{Y_{\o+1}^{(+)}(x+\frac{\ve_1}{2}+\ve_2\d_{2N-1,\o}) Y_{2N-\o-1}^{(-)}(-x-\frac{\ve_1}{2}-\ve_2) Y_{\o}^{(+)}(x-\frac{\ve_1}{2}) Y_{2N-\o}^{(-)}(-x+\frac{\ve_1}{2}+\ve_2(1-\d_{\o,0})) }
\end{split}
\end{align}
The fractional $Y$-observable obeys
\begin{align}
\begin{split}
    Y_{\o}^{(+)}(x)[\hat{\boldsymbol\lambda}] & = \BE \left[ -e^x \bS_\o^* \right] = \BE \left[ - e^x \left( \bN_\o - P_1 q_1^{-\frac{1}{2}}\bK_{\o+\frac12} + q_1^{-\frac{1}{2}} P_1 q_2^{\d_{\o,0}} \bK_{\o-\frac{1}{2}} \right)^* \right] \\
    Y_{2N-\o}^{(-)}(-x)[\hat{\boldsymbol\lambda}] & = \BE \left[ -e^{-x}\bS_{2N-\o}^* \right] = \BE \left[ - e^{-x} \left( \bN_{2N-\o} - P_1 q_1^{-\frac{1}{2}}\bK_{2N-\o-\frac12} + q_1^{-\frac{1}{2}} P_1 q_2^{\d_{\o,0}} \bK_{2N-\o+\frac{1}{2}} \right)^* \right] 
\end{split}
\end{align}
The fractional masses term are
\begin{align}\label{eq:frac-mass}
    P_\o(x) = 
    \begin{cases}
        \sh(2x+\ve_1) & \o = N-1,\ 2N-1; \\
        \sh(2x-\ve_1) & \o = 0,\ N; \\ 
        0 & \text{otherwise}.
    \end{cases}
\end{align}
The $2N$ fractional $qq$-character observables are
\begin{align}\label{eq:frac-qq}
\begin{split}
    \EX_\o (X)[\hat{\boldsymbol\lambda}] 
    & = 
    Y_{\o+1}^{(+)} \left(x+\frac{\ve_1}{2}+\ve_2\d_{\o,2N-1}\right)[\hat{\boldsymbol\lambda}] Y_{2N-\o-1}^{(-)} \left(-x-\frac{\ve_1}{2}-\ve_2\right)[\hat{\boldsymbol\lambda}] \\ 
    & \qquad + \frac{\kq_{\o+\frac{1}{2}} P_\o(x)}{Y_{\o}^{(+)}\left(x-\frac{\ve_1}{2}\right)[\hat{\boldsymbol\lambda}]  Y_{2N-\o}^{(-)}\left(-x+\frac{\ve_1}{2}+\ve_2(1-\d_{\o,0})\right)[\hat{\boldsymbol\lambda}]}.
\end{split}
\end{align}
Their expectation values
\begin{align}\label{eq:frac-qq-exp}
    \langle \EX_\o(X) \rangle = T_\o(X)
    & = \frac{1}{\hat\CalZ^{(SO)}} \sum_{\hat{\boldsymbol\lambda}} \prod_{\o=0}^{2N-1} \kq_{\o+\frac12}^{k_{\o+\frac12}^+} \hat\CalZ^{(SO)}[\hat{\boldsymbol\lambda}] \EX_\o(X) [\hat{\boldsymbol\lambda}]
\end{align}
are analytic functions in $x$ \cite{Nikita:V}. 

\subsection{Quantum Hamiltonian}
A function $f(X=e^x)$ analytic in $x$ means it can only have poles at $X=\infty$ ($x=+\infty$) and $X=0$ ($x=-\infty$)\footnote{Here we assume $f(X)$ has no branch cuts for $X$}. We can consider both large $X$ and small $X$ expansion of the function $f(X)$. The two expansion must match since they come from the same function. 

We can expand the expectation value of the fractional $qq$-characters \eqref{eq:frac-qq-exp} $T_\o(X)$ in two ways. The $X>>1$ expansion yields
\begin{align}
    T_\o(X) = \sum_{j=0}^\infty c^{(+)}_{j,\o} X^{\frac{1}{2}-j}. 
\end{align}
The small $X<<1$ expansion yields 
\begin{align}
    T_\o(X) = \sum_{j=0}^\infty c^{(-)}_{j,\o} X^{j-\frac{1}{2}}. 
\end{align}
The expansion coefficients $c^{(\pm)}_{j,\o}$ are obtained through expanding of building blocks $Y_\o^{(+)}(x) Y^{(-)}_{2N-\o}(x)$ of the fractional $qq$-characters. For $\o=0,\dots,N-1$:
\begin{align}
\begin{split}
    & Y_\o^{(+)}\left(x\right)Y_{2N-\o}^{(-)}(-x) \\ 
    & =   \left( e^{\frac{-b_{\o+\frac12}}{2}} \sqrt{X} - \frac{e^{\frac{b_{\o+\frac12}}{2}}}{\sqrt{X}} \right) 
    \prod_{\Box \in \EK^+_{\o+\frac12}} q_1^{-\frac{1}{2}} \frac{1-\frac{e^{c_\Box}q_1^{\frac12}}{X}}{ 1-\frac{e^{c_\Box}q_1^{-\frac12} }{X} } 
    \prod_{\Box \in \EK^+_{\o-\frac12}} q_1^{\frac{1}{2}} \frac{1- \frac{e^{c_\Box}q_1^{-\frac12}}{X} }{1-\frac{e^{c_\Box}q_1^{\frac12}}{X} } \\
    & \qquad \qquad \times \prod_{\Box\in \EK^-_{2N-\o+\frac12}} q_1^{\frac{1}{2}} \frac{1-\frac{e^{-c_\Box}q_1^{-\frac{1}{2}}}{X}}{1-\frac{e^{-c_\Box}q_1^{\frac12}}{X}} \prod_{\Box\in \EK^-_{2N-\o-\frac12}} q_1^{-\frac{1}{2}} \frac{1-\frac{e^{-c_\Box}q_1^{\frac12}}{X}}{1-\frac{e^{-c_\Box}q_1^{-\frac{1}{2}}}{X}}. \\
\end{split}
\end{align}
For $\o=N,\dots,2N-1$: 
\begin{align}
\begin{split}
    & Y_\o^{(+)}(x)Y_{2N-\o}^{(-)}(-x) \\ 
    & =   \left( \frac{e^{\frac{-b_{2N-\o-\frac12}}{2}}}{\sqrt{X}} - {e^{\frac{b_{2N-\o-\frac12}}{2}}}{\sqrt{X}} \right) 
    \prod_{\Box \in \EK^+_{\o+\frac12}} q_1^{-\frac{1}{2}} \frac{1-\frac{e^{c_\Box}q_1^{\frac12}}{X}}{ 1-\frac{e^{c_\Box}q_1^{-\frac12} }{X} } 
    \prod_{\Box \in \EK^+_{\o-\frac12}} q_1^{\frac{1}{2}} \frac{1- \frac{e^{c_\Box}q_1^{-\frac12}}{X} }{1-\frac{e^{c_\Box}q_1^{\frac12}}{X} } \\
    & \qquad \qquad \qquad \times \prod_{\Box\in \EK^-_{2N-\o+\frac12}} q_1^{\frac{1}{2}} \frac{1-\frac{e^{-c_\Box}q_1^{-\frac{1}{2}}}{X}}{1-\frac{e^{-c_\Box}q_1^{\frac12}}{X}} \prod_{\Box\in \EK^-_{2N-\o-\frac12}} q_1^{-\frac{1}{2}} \frac{1-\frac{e^{-c_\Box}q_1^{\frac12}}{X}}{1-\frac{e^{-c_\Box}q_1^{-\frac{1}{2}}}{X}}. \\
\end{split}
\end{align}
Let us focus on the coefficient of $X^{-\frac{1}{2}}$ term in both the large and small $X$ expansion of $T_\o(X)$. Coming from the same function requires the coefficients to match:
\begin{align}
    c^{(+)}_{1,\o} - c^{(-)}_{0,\o} = 0. 
\end{align}
Define the fractional variables
\begin{align}\label{def:frac-u}
    \kq_{\o+\frac{1}{2}} = R^{2-2\d_{\o+\frac12,N-\frac12}-2\d_{\o+\frac12,2N-\frac12} } \frac{u_{\o+\frac32}}{u_{\o+\frac12}}.
\end{align}
\[
    u_{\o+\frac12} = u_{2N+\o+\frac12}, \ \nabla^u_{\o+\frac12} = \nabla^\kq_{\o-\frac12} - \nabla^\kq_{\o+\frac{1}{2}}
\]
such that by \eqref{eq:K_w symmetry} we can write
\begin{align}
\begin{split}
    & \left\langle k^+_{\o-\frac12} - k^+_{\o+\frac12} + k^-_{2N-\o+\frac12} - k^+_{2N-\o-\frac12} \right\rangle_\BZ \hat\CalZ^{(SO)} \\
    & = \sum_{\hat{\boldsymbol\lambda}} \prod_{\o'=0}^{2N-1} \kq_{\o'+\frac12}^{k^+_{\o'+\frac12}} \hat\CalZ^{(SO)}[\hat{\boldsymbol\lambda}] \left(k^+_{\o-\frac12} - k^+_{\o+\frac12} + k^-_{2N-\o+\frac12} - k^+_{2N-\o-\frac12}\right) \\
    & = 2 \sum_{\hat{\boldsymbol\lambda}} \prod_{\o'=0}^{2N-1} \kq_{\o'+\frac12}^{k^+_{\o'+\frac12}} \hat\CalZ^{(SO)}[\hat{\boldsymbol\lambda}] \left(k^+_{\o-\frac12} - k^+_{\o+\frac12}\right) \\
    & = 2 \nabla^u_{\o+\frac{1}{2}} \hat\CalZ^{(SO)}.
\end{split} 
\end{align}
We consider linear combination 
\begin{align}
    0 = \sum_{\o=0}^{2N-1} \hat{C}_\o \left( c^{(+)}_{1,\o} - c^{(-)}_{0,\o} \right) = \hat{\rm H} \langle \psi \rangle  - \langle (\bS+\bS^*) \psi \rangle 
\end{align}
with properly chosen coefficients $\hat{C}_\o$.  
The Hamiltonian takes the form
\begin{align}\label{def:Hamiltonian}
\begin{split}
    \hat{\rm H} = &\sum_{\o=0}^{2N-1} e^{-\hat{\tp}_{\o+\frac12}} + \sum_{\o=2}^{N-2} \kq_{\o-\frac12} e^{-\frac{\hat{\tp}_{\o+\frac12}+\hat{\tp}_{\o-\frac12}}{2}} + \sum_{\o=N+2}^{2N-2} \kq_{\o-\frac12} e^{-\frac{\hat{\tp}_{\o+\frac12}+\hat{\tp}_{\o-\frac12}}{2}} \\
    & + \kq_{N+\frac12}\kq_{N-\frac12}\kq_{N-\frac32} e^{\frac{-\hat{\tp}_{N+\frac32}-2\hat{\tp}_{N+\frac12}-2\hat{\tp}_{N-\frac12}-\hat{\tp}_{N-\frac32}}{2}} + \kq_{N-\frac12} \left( e^{\frac{-\hat{\tp}_{N+\frac12}-3\hat{\tp}_{N-\frac12}}{2}} + e^{\frac{\hat{\tp}_{N-\frac12}-\hat{\tp}_{N+\frac12}}{2}} \right) \\
    & + \kq_{N-\frac{3}{2}} \left(1+\kq_{N-\frac12} e^{\frac{-\hat{\tp}_{N-\frac12}-\hat{\tp}_{N+\frac12}}{2}}\right) e^{\frac{\hat{\tp}_{N-\frac12}-\hat{\tp}_{N-\frac32}}{2}} \\
    & + \kq_{N+\frac12} \left( e^{\frac{\hat{\tp}_{N+\frac12}-\hat{\tp}_{N+\frac32}}{2}} + e^{\frac{-3\hat{\tp}_{N+\frac12}-\hat{\tp}_{N+\frac32}}{2}} \right) + \kq_{N+\frac12} \kq_{N-\frac12} \left( e^{\frac{\hat{\tp}_{N-\frac12}-2\hat{\tp}_{N+\frac12}-\hat{\tp}_{N+\frac32}}{2}} + e^{\frac{-3\hat{\tp}_{N-\frac12}-2\hat{\tp}_{N+\frac12}-\hat{\tp}_{N+\frac32}}{2}} \right) \\
    & + \kq_{\frac12}\kq_{2N-\frac12}\kq_{2N-\frac32} e^{\frac{-\hat{\tp}_{\frac32}-2\hat{\tp}_{\frac12}-2\hat{\tp}_{2N-\frac12}-\hat{\tp}_{2N-\frac32}}{2}} + \kq_{2N-\frac12} \left( e^{\frac{-\hat{\tp}_{\frac12}-3\hat{\tp}_{2N-\frac12}}{2}} + e^{\frac{\hat{\tp}_{2N-\frac12}-\hat{\tp}_{\frac12}}{2}} \right) \\
    & + \kq_{2N-\frac{3}{2}} \left(1+\kq_{2N-\frac12} e^{\frac{-\hat{\tp}_{2N-\frac12}-\hat{\tp}_{\frac12}}{2}}\right) e^{\frac{\hat{\tp}_{2N-\frac12}-\hat{\tp}_{2N-\frac32}}{2}} \\
    & + \kq_{\frac12} \left( e^{\frac{\hat{\tp}_{\frac12}-\hat{\tp}_{\frac32}}{2}} + e^{\frac{-3\hat{\tp}_{\frac12}-\hat{\tp}_{\frac32}}{2}} \right) + \kq_{\frac12} \kq_{2N-\frac12} \left( e^{\frac{\hat{\tp}_{2N-\frac12}-2\hat{\tp}_{\frac12}-\hat{\tp}_{\frac32}}{2}} + e^{\frac{-3\hat{\tp}_{2N-\frac12}-2\hat{\tp}_{2N+\frac12}-\hat{\tp}_{2N+\frac32}}{2}} \right) \\
\end{split}
\end{align}
with 
\[
    \hat\tp_{\o+\frac12} = \begin{cases}
        2\ve_1 \nabla^u_{\o+\frac12} - b_{\o+\frac12} - \ve_1 & \o=0,\dots,N-1, \\
        2\ve_1 \nabla^u_{\o+\frac12} + b_{2N-\o-\frac12} + \ve_1 & \o=N,\dots,2N-1.
    \end{cases}
\] 
See Appendix.~\ref{sec:detail-qq} for calculation detail. 

In the NS-limit $\ve_2 \to 0$, $\ve_1\equiv \hbar$, the bulk instanton partition function $\CalZ^{(SO)}$ is dominated by the limit shape configuration $\boldsymbol\Lambda$ \eqref{def:limit-shape}. The vev of the surface defect becomes
\begin{align}
\begin{split}
    & \lim_{\ve_2 \to 0} \langle \psi \rangle = \psi[\boldsymbol\Lambda] :=\psi(\bb,\ve_1;u_\o;\kq) \\ 
    & \lim_{\ve_2\to0} \langle (\bS+\bS^*) \psi \rangle = (\bS+\bS^*)[\boldsymbol\Lambda] \psi (\bb,\ve_1;u_\o;\kq) 
\end{split}
\end{align}
The Dyson-Schwinger equation becomes Schr\"{o}dinger equation
\begin{align}
    \hat{\rm H} \psi (\bb,\ve_1;u_\o;\kq) = (\bS+\bS^*)[\boldsymbol\Lambda](\bb,\ve_1,\kq) \psi (\bb,\ve_1;u_\o;\kq). 
\end{align}

We now claim that the Hamiltonian $\hat{\rm H}$ in \eqref{def:Hamiltonian} is equivalent to $\hat{D}_N$ RTL Hamiltonian after folding and change of variables. 
This can be better illustrated on the classical level.  
First, we consider shifting
\begin{align}\label{eq:cov-1}
\begin{split}
    & u_{N-\frac{1}{2}} \to u_{N-\frac12} \sqrt{e^{2\tp_{N-\frac12}}+e^{-2\tp_{N-\frac12}}}, \quad u_{N+\frac{1}{2}} \to u_{N+\frac12} \sqrt{e^{2\tp_{N+\frac12}}+e^{-2\tp_{N+\frac12}}}, \\
    & u_{2N-\frac{1}{2}} \to u_{2N-\frac12} \sqrt{e^{2\tp_{2N-\frac12}}+e^{-2\tp_{2N-\frac12}}}, \quad u_{\frac{1}{2}} \to u_{\frac12} \sqrt{e^{2\tp_{\frac12}}+e^{-2\tp_{\frac12}}}.
\end{split}
\end{align}
Secondly we reduce the degree of freedom of the system by half through folding:
\[
    u_{\o+\frac{1}{2}} = u^{-1}_{2N-\o-\frac{1}{2}}, \quad \tp_{\o+\frac{1}{2}} = - \tp_{2N-\o-\frac{1}{2}}
\]
Finally we take another change of variables on the reduced variables
\begin{align}\label{eq:cov-2}
\begin{split}
    & e^{2\tq_{N-\frac12}} \to \frac{\cosh \tq_{N-\frac12}}{\cosh \frac{\tp_{N-\frac12}}{2}}, \quad e^{2\tp_{N-\frac12}} \to \frac{\cosh\frac{\tp_{N-\frac12}-2\tq_{N-\frac12}}{2}}{\cosh\frac{\tp_{N-\frac12}+2\tq_{N-\frac12}}{2}}, \\
    & e^{-2\tq_{\frac12}} \to \frac{\cosh \tq_{\frac12}}{\cosh \frac{\tp_{\frac12}}{2}}, \quad e^{-2\tp_{\frac12}} \to \frac{\cosh\frac{\tp_{\frac12}-2\tq_{\frac12}}{2}}{\cosh\frac{\tp_{\frac12}+2\tq_{\frac12}}{2}}. 
\end{split}  
\end{align}
The Hamiltonian \eqref{def:Hamiltonian} recovers the $\hat{D}_N$ RTL Hamiltonian \eqref{def:H-D-RTL}: 
\begin{align}
\begin{split}
    \hat{\rm H} = &\sum_{\o=0}^{N-1} 2\cosh \hat{\tp}_{\o+\frac12} + \sum_{\o=1}^{N-1} 2 R^2 e^{\tq_{\o+\frac{1}{2}}-\tq_{\o-\frac{1}{2}}} \cosh \frac{\hat{\tp}_{\o+\frac12}+\hat{\tp}_{\o-\frac12}}{2} \\
    &  + R^4 e^{2\tq_{\frac32}} + 2R^2 e^{\tq_{\frac12}+\tq_{\frac32}} \cosh\frac{\hat{\tp}_{\frac12}-\hat{\tp}_{\frac32} }{2} \\
     & + R^4 e^{-2\tq_{N-\frac32}} + 2R^2 e^{-\tq_{N-\frac12}-\tq_{N-\frac32}} \cosh\frac{\hat{\tp}_{N-\frac12}-\hat{\tp}_{N-\frac32}}{2} 
\end{split}
\end{align}
with $\log u_{\o+\frac12} = {\tq_{\o+\frac12}}$.

\section{Summery and future direction}\label{sec:summary}

In this article we explore the Bethe/Gauge correspondence of type D relativistic Toda lattice and its gauge theory dual. 
Due to lack of Young diagram description for $SO(2N)$ supersymmetry gauge theory instanton partition function, we consider $U(2N)$ gauge theory with eight fundamental matters which shares the same toric diagram with the $SO(2N)$ theory.
The fundamental masses and Coulomb moduli of the $U(2N)$ gauge theory are fine-tuned to allow folding from $U(2N)$ gauge theory to $SO(2N)$ theory.

We study the instanton moduli space of $U(2N)$ group with a modified real and complex moment map \eqref{def:SU-SO-moduli}. A class of solution of the modified real and complex moment mapa has the ADHM matrices $B_{1}$ and $B_2$ be block diagonal. 
We prove the stability condition for the ADHM vector spaces \eqref{eq:stability-folding} in the said instanton moduli space. The instanton configuration can be represented by a set of $N$-tuples of Young diagrams. 
In the unrefined limit $\ve_+ = 0$ our result matches with the JK-residue computation \cite{Nawata:2021dlk}. 
In the NS-limit $\ve_2 \to 0$ the twisted superpotential \eqref{def:twist-super-potential} matches with the effective potential of the 3d $\CalN=2$ theory \cite{Kimura:2020bed,Ding:2023lsk}. 

The correspondence between the type D RTL and the folded gauge theory is established on two levels: 
First, we recover the Bethe ansatz equation of type D RTL from the $qq$-character of the gauge theory. 
Secondly, by introducing co-dimensional two orbifold defect, we recover the $\hat{D}_N$ RTL Hamiltonian from the fractional $qq$-character. The defect partition function in the NS-limit is proven as the wavefunction of the $\hat{D}_N$ RTL. 

Let us end this note with a few remarks and furture directions

\begin{itemize}
    \item We would like to study the potential folding of ADHM construction from type A to type B and C. In particular we are curious whether a proper modification of the real and complex moment maps can lead to Young diagram description of the instanton configuration. 
    \item Two change of variables \eqref{eq:cov-1} and \eqref{eq:cov-2} are performed in order to fully match the $\hat{D}_N$ RTL Hamiltonian with what we obtained from the fractional $qq$-characters. 
    A similar but not identical change of variable is also observed in the study of dimer graphs for type $D$ RTL \cite{Lee:2024bqg}. 
    Despite being straightforward in the classical level, the meaning of these change of variables \eqref{eq:cov-1} and \eqref{eq:cov-2} in the quantum level is not well understood at this moment. 
    \item Finally, we would like to know if the Bethe/Gauge correspondence can be uplifted to 6d $\CalN=(1,0)$ theory. The R-matrix of the elliptic integrable system is associated to the elliptic Ding-Iohara-Miki algebra \cite{saito2014elliptic}. 
\end{itemize}

\newpage
\appendix

\section{Instanton pseudo measure}

The pseudo-measure of SO gauge group:
\begin{align}
\begin{split}
    & \frac{d\phi_I}{2\pi\ri} \prod_{I} \frac{\sh\ve_+}{\sh\ve_1 \sh\ve_2} \frac{\sh(2\phi_I)^2\sh(2\phi_I+\ve_+)\sh(2\phi_I-\ve_+)}{\prod_{\beta=1}^N \sh \left( \frac{\ve_+}{2} \pm \phi_I \pm b_\beta \right)} \\
    & \times \prod_{I\neq J} \frac{\sh(\phi_I-\phi_J)\sh(\phi_I-\phi_J+\ve_+)}{\sh(\phi_I-\phi_J+\ve_1)\sh(\phi_I-\phi_J+\ve_2)} \times \prod_{I<J} \frac{\sh\pm(\phi_I+\phi_J)\sh(\ve_+\pm (\phi_I+\phi_J))}{\sh(\ve_1\pm (\phi_I+\phi_J))\sh(\ve_2\pm (\phi_I+\phi_J))}
\end{split}
\end{align}
Can be rewritten into $\BE$-functor by
\begin{align}
\begin{split}
    & \BE \left[ -\frac{(\bS+\bS^*)^2}{2P_{12}^*} \right] \\
    & = \BE \left[ -\frac{(\bN+\bN^* + q_+^{-\frac{1}{2}}P_{12}\bK + q_+^{\frac{1}{2}}P_{12}^*\bK^* )^2}{2P_{12}^*} \right] \\
    & = \BE \left[ -\frac{(\bN+\bN^*)^2}{2P_{12}^*} \right] \BE \left[ q_+^{\frac{1}{2}} (\bN+\bN^*)(\bK+\bK^*) \right] \BE \left[ -\frac{1}{2}P_{12}(\bK+\bK^*)^2 \right]
\end{split}
\end{align}
where
\begin{align}
\begin{split}
    & \BE \left[ - \frac{1}{2} P_{12} (\bK+\bK^*) (\bK+\bK^*) \right] \\
    & = \BE \left[ -\frac{1}{2} P_{12} \left( \sum_{I=1}^k e^{\phi_I} + e^{-\phi_I} \right) \left( \sum_{J=1}^k e^{\phi_J} + e^{-\phi_J} \right) \right] \\
    & = \BE \left[ -\frac{1}{2} P_{12} \sum_{I<J} e^{\phi_I + \phi_J} + e^{\phi_I-\phi_J} + e^{\phi_J-\phi_I} + e^{-\phi_I-\phi_J} \right] \\
    & \quad \times \BE \left[ -\frac{1}{2} P_{12} \sum_{I>J} e^{\phi_I + \phi_J} + e^{\phi_I-\phi_J} + e^{\phi_J-\phi_I} + e^{-\phi_I-\phi_J} \right] \\
    & \quad \times \BE \left[ - \frac{1}{2} P_{12} \sum_{I} e^{2\phi_I} + 2 + e^{-2\phi_I} \right] \\
    & = \BE \left[ -P_{12}\sum_{I\neq J} e^{\phi_I-\phi_J} \right] \BE \left[ - P_{12} \sum_{I<J} e^{\phi_I+\phi_J} + e^{-\phi_I-\phi_J} \right] \BE \left[ -\frac{1}{2}P_{12} \sum_{I} e^{2\phi_I} +2+ e^{-2\phi_I} \right] \\
    & = \prod_{I\neq J} \frac{\sh(\phi_I-\phi_J)\sh(\phi_I-\phi_J+\ve_+)}{\sh(\phi_I-\phi_J+\ve_1)\sh(\phi_I-\phi_J+\ve_2)} \times \prod_{I<J} \frac{\sh\pm(\phi_I+\phi_J)\sh(\ve_+\pm (\phi_I+\phi_J))}{\sh(\ve_1\pm (\phi_I+\phi_J))\sh(\ve_2\pm (\phi_I+\phi_J))} \\
    & \qquad \times \frac{\sh\ve_+}{\sh\ve_1\sh\ve_2} \times \left[ \frac{\sh(2\phi_I)\sh(-2\phi_I)\sh(\ve_++2\phi_I)\sh(\ve_+-2\phi_I)}{\sh(\ve_1+2\phi_I)\sh(\ve_1-2\phi_I)\sh(\ve_2+2\phi_I)\sh(\ve_2-2\phi_I)}  \right]^{\frac{1}{2}}
\end{split}
\end{align}
The contribution from the fundamental masses in $U(2N)$ is
\begin{align}
\begin{split}
    & \BE \left[ -\bM (\bK+\bK^*) \right] \\
    & = \BE \left[ - \sum_{f=1}^8 \sum_I e^{m_f+\phi_I} + e^{m_f-\phi_I} \right] \\
    & = \prod_{f=1}^8 \prod_{I} \sh(\phi_I+m_f)\sh(m_f-\phi_I)
\end{split}
\end{align}
Let us consider the case 
\[
    m_{f+4} = m_f + \ri\pi
\]
The mass contribution becomes
\begin{align}
\begin{split}
    & \prod_{f=1}^4 \prod_{I} \sh(\phi_I+m_f)\sh(m_f-\phi_I) \sh(\phi_I+m_f+\ri\pi)\sh(m_f+\ri\pi-\phi_I) \\
    & = \prod_{f=1}^4 \prod_I \sh(2\phi_I+2m_f) \sh(2m_f-2\phi_I) \\
    & = \BI \left[ -\bM' (\bK+\bK^*) \right], \ \bM' = \sum_{f=1}^4 e^{m_f}
\end{split}
\end{align}
To match with the $SO(2N)$ measure, we choose 
\[
    m_1 = 0, \ m_2 = \frac{\ve_1}{2}, \ m_3 = \frac{\ve_2}{2}, \ m_4 = \frac{\ve_+}{2}
\]

\section{Detail calculation of \textit{qq}-character} \label{sec:detail-qq}

The fractional $qq$-characters \eqref{eq:frac-qq}
\begin{align}
\begin{split}
    \EX_\o (X)[\hat{\boldsymbol\lambda}] 
    & = 
    Y_{\o+1}^{(+)} \left(x+\frac{\ve_1}{2}+\ve_2\d_{\o,2N-1}\right)[\hat{\boldsymbol\lambda}]  Y_{2N-\o-1}^{(-)} \left(-x-\frac{\ve_1}{2}-\ve_2\right)[\hat{\boldsymbol\lambda}] \\ 
    & \qquad + \frac{\kq_{\o+\frac{1}{2}} P_\o(x)}{Y_{\o}^{(+)}\left(x-\frac{\ve_1}{2}\right)[\hat{\boldsymbol\lambda}] Y_{2N-\o}^{(-)}\left(-x+\frac{\ve_1}{2}+\ve_2(1-\d_{\o,0})\right)[\hat{\boldsymbol\lambda}]}
\end{split}
\end{align}
is build by fractional $Y_\o^{(\pm)}(x)$. $P_\o(x)$ is defined in \eqref{eq:frac-mass}.
For $\o=0,\dots,N-1$:
\begin{align}
\begin{split}
    & Y_\o^{(+)}\left(x\right)Y_{2N-\o}^{(-)}(-x) \\ 
    & = \BE \left[ -X (\bN_{\o+\frac12} - P_1 q_1^{-\frac12} \bK_{+,\o+\frac12} + P_1 q_1^{-\frac{1}{2}}  \bK_{+,{\o-\frac12}} )^* + X^* \left( P_1 q_1^{-\frac{1}{2}} \bK_{-,2N-\o+\frac{1}{2}} - P_1q_1^{-\frac12} \bK_{-,2N-\o-\frac12} \right)^* \right] \\
    & =   \left( e^{\frac{-b_{\o+\frac12}}{2}} \sqrt{X} - \frac{e^{\frac{b_{\o+\frac12}}{2}}}{\sqrt{X}} \right) 
    \prod_{\Box \in \EK^+_{\o+\frac12}} q_1^{-\frac{1}{2}} \frac{1-\frac{e^{c_\Box}q_1^{\frac12}}{X}}{ 1-\frac{e^{c_\Box}q_1^{-\frac12} }{X} } 
    \prod_{\Box \in \EK^+_{\o-\frac12}} q_1^{\frac{1}{2}} \frac{1- \frac{e^{c_\Box}q_1^{-\frac12}}{X} }{1-\frac{e^{c_\Box}q_1^{\frac12}}{X} } \\
    & \qquad \times \prod_{\Box\in \EK^-_{2N-\o+\frac12}} q_1^{\frac{1}{2}} \frac{1-\frac{e^{-c_\Box}q_1^{-\frac{1}{2}}}{X}}{1-\frac{e^{-c_\Box}q_1^{\frac12}}{X}} \prod_{\Box\in \EK^-_{2N-\o-\frac12}} q_1^{-\frac{1}{2}} \frac{1-\frac{e^{-c_\Box}q_1^{\frac12}}{X}}{1-\frac{e^{-c_\Box}q_1^{-\frac{1}{2}}}{X}}. \\
\end{split}
\end{align}
For $\o=N,\dots,2N-1$: 
\begin{align}
\begin{split}
    & Y_\o^{(+)}(x)Y_{2N-\o}^{(-)}(-x) \\ 
    & = \BE \left[ X ( P_1 q_1^{-\frac12} \bK_{+,\o+\frac12} - P_1 q_1^{-\frac{1}{2}} \bK_{{+,\o-\frac12}} )^* + X^* \left( - \bN_{2N-\o-\frac12}^* + P_1 q_1^{-\frac{1}{2}} \bK_{-,2N-\o+\frac{1}{2}} - P_1q_1^{-\frac12} \bK_{-,2N-\o-\frac12} \right)^* \right] \\
    & =   \left( \frac{e^{\frac{-b_{2N-\o-\frac12}}{2}}}{\sqrt{X}} - {e^{\frac{b_{2N-\o-\frac12}}{2}}}{\sqrt{X}} \right) 
    \prod_{\Box \in \EK^+_{\o+\frac12}} q_1^{-\frac{1}{2}} \frac{1-\frac{e^{c_\Box}q_1^{\frac12}}{X}}{ 1-\frac{e^{c_\Box}q_1^{-\frac12} }{X} } 
    \prod_{\Box \in \EK^+_{\o-\frac12}} q_1^{\frac{1}{2}} \frac{1- \frac{e^{c_\Box}q_1^{-\frac12}}{X} }{1-\frac{e^{c_\Box}q_1^{\frac12}}{X} } \\
    & \qquad \times \prod_{\Box\in \EK^-_{2N-\o+\frac12}} q_1^{\frac{1}{2}} \frac{1-\frac{e^{-c_\Box}q_1^{-\frac{1}{2}}}{X}}{1-\frac{e^{-c_\Box}q_1^{\frac12}}{X}} \prod_{\Box\in \EK^-_{2N-\o-\frac12}} q_1^{-\frac{1}{2}} \frac{1-\frac{e^{-c_\Box}q_1^{\frac12}}{X}}{1-\frac{e^{-c_\Box}q_1^{-\frac{1}{2}}}{X}}. \\
\end{split}
\end{align}

The expectation value of $qq$-character can be calculated based on the building block. In the case $\o=1,\dots,N-2$: 
\begin{align}
\begin{split}
    \EX_\o(X)
    & = \left( e^{\frac{-b_{\o+\frac32}}{2}}q_1^{\frac{1}{4}} \sqrt{X} - \frac{e^{\frac{b_{\o+\frac32}}{2}}q_1^{-\frac14}}{\sqrt{X}} \right) 
    \prod_{\Box \in \EK^+_{\o+\frac32}} q_1^{-\frac{1}{2}} \frac{1-\frac{e^{c_\Box}}{X}}{ 1-\frac{e^{c_\Box}q_1^{-1} }{X} } 
    \prod_{\Box \in \EK^+_{\o+\frac12}} q_1^{\frac{1}{2}} \frac{1- \frac{e^{c_\Box}q_1^{-1}}{X} }{1-\frac{e^{c_\Box}}{X} } \\
    & \qquad \times \prod_{\Box\in \EK^-_{2N-\o-\frac12}} q_1^{\frac{1}{2}} \frac{1-\frac{e^{-c_\Box}q_1^{-1}}{X}}{1-\frac{e^{-c_\Box}}{X}} \prod_{\Box\in \EK^-_{2N-\o-\frac32}} q_1^{-\frac{1}{2}} \frac{1-\frac{e^{-c_\Box}}{X}}{1-\frac{e^{-c_\Box}q_1^{-1}}{X}}. \\
    & + \frac{\kq_{\o+\frac{1}{2}} }{e^{\frac{-b_{\o+\frac12}}{2}} q_1^{-\frac{1}{4}} \sqrt{X} - \frac{e^{\frac{b_{\o+\frac12}}{2}}q_1^{\frac14}}{\sqrt{X}}} 
    \prod_{\Box \in \EK^+_{\o+\frac12}} q_1^{\frac{1}{2}} \frac{ 1-\frac{e^{c_\Box} }{X} }{1-\frac{e^{c_\Box}q_1}{X}} 
    \prod_{\Box \in \EK^+_{\o-\frac12}} q_1^{-\frac{1}{2}} \frac{1-\frac{e^{c_\Box}q_1}{X} }{1- \frac{e^{c_\Box}}{X} } \\
    & \qquad \times \prod_{\Box\in \EK^-_{2N-\o+\frac12}} q_1^{-\frac{1}{2}} \frac{1-\frac{e^{-c_\Box}q_1}{X}}{1-\frac{e^{-c_\Box}}{X}} \prod_{\Box\in \EK^-_{2N-\o-\frac12}} q_1^{\frac{1}{2}} \frac{1-\frac{e^{-c_\Box}}{X}}{1-\frac{e^{-c_\Box}q_1}{X}}. \\
\end{split}
\end{align}
Similarly for $\o=N+1,\dots,2N-2$ 
\begin{align}
\begin{split}
    \EX_\o(X)
    & = \left( \frac{e^{\frac{-b_{2N-\o-\frac32}}{2}}q_1^{-\frac{1}{4}}}{\sqrt{X}} - {e^{\frac{b_{2N-\o-\frac32}}{2}}}q_1^{\frac{1}{4}}{\sqrt{X}} \right)  
    \prod_{\Box \in \EK_{\o+\frac32}^+} q_1^{-\frac{1}{2}} \frac{1-\frac{e^{c_\Box}}{X}}{ 1-\frac{e^{c_\Box}q_1^{-1} }{X} } 
    \prod_{\Box \in \EK_{\o+\frac12}^+} q_1^{\frac{1}{2}} \frac{1- \frac{e^{c_\Box}q_1^{-1}}{X} }{1-\frac{e^{c_\Box}}{X} } \\
    & \qquad \times \prod_{\Box\in \EK_{2N-\o-\frac12}^-} q_1^{\frac{1}{2}} \frac{1-\frac{e^{-c_\Box}q_1^{-1}}{X}}{1-\frac{e^{-c_\Box}}{X}} \prod_{\Box\in \EK_{2N-\o-\frac32}^-} q_1^{-\frac{1}{2}} \frac{1-\frac{e^{-c_\Box}}{X}}{1-\frac{e^{-c_\Box}q_1^{-1}}{X}}. \\
    & + \frac{\kq_{\o+\frac{1}{2}} }{\frac{e^{\frac{-b_{2N-\o-\frac12}}{2}}q_1^{\frac{1}{4}}}{\sqrt{X}} - {e^{\frac{b_{2N-\o-\frac12}}{2}}}q_1^{-\frac{1}{4}}{\sqrt{X}} } 
    \prod_{\Box \in \EK_{\o+\frac12}^+} q_1^{\frac{1}{2}} \frac{ 1-\frac{e^{c_\Box} }{X} }{1-\frac{e^{c_\Box}q_1}{X}} 
    \prod_{\Box \in \EK_{\o-\frac12}^+} q_1^{-\frac{1}{2}} \frac{1-\frac{e^{c_\Box}q_1}{X} }{1- \frac{e^{c_\Box}}{X} } \\
    & \qquad \times \prod_{\Box\in \EK_{2N-\o+\frac12}^-} q_1^{-\frac{1}{2}} \frac{1-\frac{e^{-c_\Box}q_1}{X}}{1-\frac{e^{-c_\Box}}{X}} \prod_{\Box\in \EK_{2N-\o-\frac12}^-} q_1^{\frac{1}{2}} \frac{1-\frac{e^{-c_\Box}}{X}}{1-\frac{e^{-c_\Box}q_1}{X}}. \\
\end{split}
\end{align}
For the case $\o=0$: 
\begin{align}
\begin{split}
    \EX_0(X) =
    & \left( e^{\frac{-b_{\frac32}}{2}}q_1^{\frac{1}{4}} \sqrt{X} - \frac{e^{\frac{b_{\frac32}}{2}}q_1^{-\frac14}}{\sqrt{X}} \right) 
    \prod_{\Box \in \EK^+_{\frac32}} q_1^{-\frac{1}{2}} \frac{1-\frac{e^{c_\Box}}{X}}{ 1-\frac{e^{c_\Box}q_1^{-1} }{X} } 
    \prod_{\Box \in \EK^+_{\frac12}} q_1^{\frac{1}{2}} \frac{1- \frac{e^{c_\Box}q_1^{-1}}{X} }{1-\frac{e^{c_\Box}}{X} } \\
    & \qquad \times \prod_{\Box\in \EK^-_{-\frac12}} q_1^{\frac{1}{2}} \frac{1-\frac{e^{-c_\Box}q_1^{-1}}{X}}{1-\frac{e^{-c_\Box}}{X}} \prod_{\Box\in \EK^-_{-\frac32}} q_1^{-\frac{1}{2}} \frac{1-\frac{e^{-c_\Box}}{X}}{1-\frac{e^{-c_\Box}q_1^{-1}}{X}}. \\
    & + \kq_{\frac{1}{2}}\frac{ X q_1^{-\frac{1}{2}} - \frac{q_1^{\frac12}}{X} }{e^{\frac{-b_{\frac12}}{2}} q_1^{-\frac{1}{4}} \sqrt{X} - \frac{e^{\frac{b_{\frac12}}{2}}q_1^{\frac14}}{\sqrt{X}}} 
    \prod_{\Box \in \EK^+_{\frac12}} q_1^{\frac{1}{2}} \frac{ 1-\frac{e^{c_\Box} }{X} }{1-\frac{e^{c_\Box}q_1}{X}} 
    \prod_{\Box \in \EK^+_{-\frac12}} q_1^{-\frac{1}{2}} \frac{1-\frac{e^{c_\Box}q_1}{X} }{1- \frac{e^{c_\Box}}{X} } \\
    & \qquad \times \prod_{\Box\in \EK^-_{\frac12}} q_1^{-\frac{1}{2}} \frac{1-\frac{e^{-c_\Box}q_1}{X}}{1-\frac{e^{-c_\Box}}{X}} \prod_{\Box\in \EK^-_{-\frac12}} q_1^{\frac{1}{2}} \frac{1-\frac{e^{-c_\Box}}{X}}{1-\frac{e^{-c_\Box}q_1}{X}}. \\
\end{split}
\end{align}
For the case $\o=N$: 
\begin{align}
\begin{split}
    \EX_{N}(X)
    & = \left( \frac{e^{\frac{-b_{N-\frac32}}{2}}q_1^{\frac{1}{4}}}{\sqrt{X}} - {e^{\frac{b_{N-\frac32}}{2}}q_1^{-\frac14}}{\sqrt{X}} \right) 
    \prod_{\Box \in \EK_{N+\frac32}^+} q_1^{-\frac{1}{2}} \frac{1-\frac{e^{c_\Box}}{X}}{ 1-\frac{e^{c_\Box}q_1^{-1} }{X} } 
    \prod_{\Box \in \EK_{N+\frac12}^+} q_1^{\frac{1}{2}} \frac{1- \frac{e^{c_\Box}q_1^{-1}}{X} }{1-\frac{e^{c_\Box}}{X} } \\
    & \qquad \times \prod_{\Box\in \EK_{N-\frac12}^-} q_1^{\frac{1}{2}} \frac{1-\frac{e^{-c_\Box}q_1^{-1}}{X}}{1-\frac{e^{-c_\Box}}{X}} \prod_{\Box\in \EK_{N-\frac32}^-} q_1^{-\frac{1}{2}} \frac{1-\frac{e^{-c_\Box}}{X}}{1-\frac{e^{-c_\Box}q_1^{-1}}{X}}. \\
    & + \kq_{N+\frac{1}{2}} \frac{ X q_1^{\frac12}  - \frac{q_1^{-\frac12}}{X} }{ \frac{e^{\frac{-b_{N-\frac12}}{2}} q_1^{-\frac{1}{4}}}{\sqrt{X}} - {e^{\frac{b_{N-\frac12}}{2}}q_1^{\frac14}}{\sqrt{X}}} 
    \prod_{\Box \in \EK^+_{N+\frac12}} q_1^{\frac{1}{2}} \frac{ 1-\frac{e^{c_\Box} }{X} }{1-\frac{e^{c_\Box}q_1}{X}} 
    \prod_{\Box \in \EK^+_{N-\frac12}} q_1^{-\frac{1}{2}} \frac{1-\frac{e^{c_\Box}q_1}{X} }{1- \frac{e^{c_\Box}}{X} } \\
    & \qquad \times \prod_{\Box\in \EK^-_{N+\frac12}} q_1^{-\frac{1}{2}} \frac{1-\frac{e^{-c_\Box}q_1}{X}}{1-\frac{e^{-c_\Box}}{X}} \prod_{\Box\in \EK^-_{N-\frac12}} q_1^{\frac{1}{2}} \frac{1-\frac{e^{-c_\Box}}{X}}{1-\frac{e^{-c_\Box}q_1}{X}}. \\
\end{split}
\end{align}
For the case $\o=N-1$: 
\begin{align}
\begin{split}
    \EX_{N-1}(X) 
    & = \left( \frac{e^{-\frac{b_{N-\frac12}}{2}} q_1^{-\frac14} }{\sqrt{X}} - e^{\frac{b_{N-\frac12}}{2}}q_1^{\frac14} \sqrt{X} \right) \prod_{\Box \in \EK_{N+\frac12}^+} q_1^{-\frac12} \frac{1-\frac{e^{c_\Box}}{X}}{ 1-\frac{e^{c_\Box}q_1^{-1} }{X} } 
    \prod_{\Box \in \EK_{N-\frac12}^+} q_1^{\frac12} \frac{1- \frac{e^{c_\Box}q_1^{-1}}{X} }{1-\frac{e^{c_\Box}}{X} } \\
    & \qquad \times \prod_{\Box \in \EK_{N+\frac12}^-} q_1^{\frac12} \frac{1-\frac{e^{-c_\Box}q_1^{-1}}{X}}{1-\frac{e^{-c_\Box}}{X}}
    \prod_{\Box \in \EK_{N-\frac12}^-} q_1^{-\frac12} \frac{1-\frac{e^{-c_\Box}}{X}}{1-\frac{e^{-c_\Box}q_1^{-1}}{X}}\\
    & \quad + \kq_{N-\frac{1}{2}}  \frac{Xq_1^{-\frac12}-\frac{q_1^{\frac12}}{X}}{e^{\frac{-b_{N-\frac12}}{2}} q_1^{-\frac{1}{4}} \sqrt{X} - \frac{e^{\frac{b_{N-\frac12}}{2}}q_1^{\frac14}}{\sqrt{X}}} 
    \prod_{\Box \in \EK_{N-\frac12}^+} q_1^{\frac{1}{2}} \frac{ 1-\frac{e^{c_\Box} }{X} }{1-\frac{e^{c_\Box}q_1}{X}} 
    \prod_{\Box \in \EK_{N-\frac32}^+} q_1^{-\frac{1}{2}} \frac{1-\frac{e^{c_\Box}q_1}{X} }{1- \frac{e^{c_\Box}}{X} } \\
    & \qquad \times \prod_{\Box\in \EK_{N+\frac32}^-} q_1^{-\frac{1}{2}} \frac{1-\frac{e^{-c_\Box}q_1}{X}}{1-\frac{e^{-c_\Box}}{X}} \prod_{\Box\in \EK_{N+\frac12}^-} q_1^{\frac{1}{2}} \frac{1-\frac{e^{-c_\Box}}{X}}{1-\frac{e^{-c_\Box}q_1}{X}}. \\
\end{split}
\end{align}
Last but not least the $\o=2N-1$ case: 
\begin{align}
\begin{split}
    & \EX_{2N-1}(X) \\
    =& \ \left( e^{\frac{-b_{\frac12} }{2}} q_1^{\frac{1}{4}} \sqrt{X} - \frac{e^{\frac{b_{\frac12} }{2}} q_1^{-\frac{1}{4}}}{\sqrt{X}} \right)  
    \prod_{\Box \in \EK_{\frac12}^+} q_1^{-\frac12} \frac{1-\frac{e^{c_\Box}}{X}}{ 1-\frac{e^{c_\Box}q_1^{-1} }{X} } 
    \prod_{\Box \in \EK_{-\frac12}^+} q_1^{\frac12} \frac{1- \frac{e^{c_\Box}q_1^{-1}}{X} }{1-\frac{e^{c_\Box}}{X} }  \\
    & \qquad \times \prod_{\Box \in \EK_{\frac12}^-} q_1^{\frac12} \frac{1-\frac{e^{-c_\Box}q_1^{-1}}{X}}{1-\frac{e^{-c_\Box}}{X}}  
    \prod_{\Box \in \EK_{-\frac12}^-} q_1^{-\frac12} \frac{1-\frac{e^{-c_\Box}}{X}}{1-\frac{e^{-c_\Box}q_1^{-1}}{X}} \\
    & + \kq_{2N-\frac{1}{2}} \frac{Xq_1^{\frac12}-\frac{q_1^{-\frac12}}{X}}{ \frac{e^{\frac{-b_{\frac12}}{2}} q_1^{\frac{1}{4}}}{\sqrt{X}} - {e^{\frac{b_{\frac12}}{2}}q_1^{-\frac14}}{\sqrt{X}}} 
    \prod_{\Box \in \EK_{-\frac12}^+} q_1^{\frac{1}{2}} \frac{ 1-\frac{e^{c_\Box} }{X} }{1-\frac{e^{c_\Box}q_1}{X}} 
    \prod_{\Box \in \EK_{-\frac32}^+} q_1^{-\frac{1}{2}} \frac{1-\frac{e^{c_\Box}q_1}{X} }{1- \frac{e^{c_\Box}q_1^{-1}}{X} } \\
    & \qquad \times \prod_{\Box\in \EK_{\frac32}^-} q_1^{-\frac{1}{2}} \frac{1-\frac{e^{-c_\Box}q_1}{X}}{1-\frac{e^{-c_\Box}}{X}} \prod_{\Box\in \EK_{\frac12}^-} q_1^{\frac{1}{2}} \frac{1-\frac{e^{-c_\Box}}{X}}{1-\frac{e^{-c_\Box}q_1}{X}}. \\
\end{split}
\end{align}
\eqref{eq:K_w symmetry} implies $k_{\o+\frac12}^+ = k^-_{2N-\o-\frac12}$. The expectation value of the fractional instanton counting $k_{\o+\frac12}^+$ can be rewrite as differential operator acting on the fractional instanton partition function
\[
    \langle k_{\o+\frac12}^+ \rangle \hat\CalZ^{(SO)} = \nabla^\kq_{\o+\frac12} \hat\CalZ^{(SO))}. 
\]

For $\o=1,\dots,N-2$, large $X$ expansion: 
\begin{align}
\begin{split}
    & \left[ X^{-\frac{1}{2}} \right] T_\o(X) \\
    & = \left \langle e^{-\frac{b_{\o+\frac32} }{2}} q_1^{\frac{1}{4}} q_1^{\frac{1}{2} (k^+_{\o+\frac12}-k^+_{\o+\frac32}+k_{2N-\o-\frac12}^--k^-_{2N-\o-\frac32}) } \right. \\
    & \qquad \times \left[ -e^{b_{\o+\frac{3}{2}} }q_1^{-\frac12} + (q_1^{-1}-1) \left( \bK_{+,\o+\frac{3}{2}} - \bK_{+,\o+\frac12} - \bK_{-,2N-\o-\frac12}^* + \bK_{-,2N-\o-\frac32}^* \right) \right] \\
    & \qquad \left. + \kq_{\o+\frac{1}{2}} e^{\frac{b_{\o+\frac12} }{2}} q_1^{\frac14} q_1^{\frac{1}{2}(k_{\o+\frac12}^+ - k_{\o-\frac12}^+ - k_{2N-\o+\frac12}^- + k_{2N-\o-\frac12}^- ) } \right \rangle 
\end{split}
\end{align}
DS equation: 
\begin{align}
\begin{split}
    0 & = \left \langle e^{-\frac{b_{\o+\frac32} }{2}} q_1^{\frac{1}{4}} q_1^{\frac{1}{2} (k^+_{\o+\frac12}-k^+_{\o+\frac32}+k_{2N-\o-\frac12}^--k^-_{2N-\o-\frac32}) } \right. \\
    & \qquad \times \left[ -e^{b_{\o+\frac{3}{2}} }q_1^{-\frac12} + (q_1^{-1}-1) \left( \bK_{+,\o+\frac{3}{2}} - \bK_{+,\o+\frac12} - \bK_{-,2N-\o-\frac12}^* + \bK_{-,2N-\o-\frac32}^* \right) \right] \\
    & \qquad + \kq_{\o+\frac{1}{2}} e^{\frac{b_{\o+\frac12} }{2}} q_1^{\frac14} q_1^{\frac{1}{2}(k_{\o+\frac12}^+ - k_{\o-\frac12}^+ - k_{2N-\o+\frac12}^- + k_{2N-\o-\frac12}^- ) } \\
    & \qquad \left. + {e^{\frac{b_{\o+\frac32}}{2}}q_1^{-\frac14}} q_1^{\frac{-1}{2}(k^+_{\o+\frac12}-k^+_{\o+\frac32}+k^-_{2N-\o-\frac12} - k^-_{2N-\o-\frac32})}\right \rangle 
\end{split}
\end{align}
For $\o=N+1,\dots,2N-2$, large $X$ expansion: 
\begin{align}
\begin{split}
    & -\left[ X^{-\frac{1}{2}} \right] T_\o(X) \\
    & = \left \langle e^{\frac{b_{2N-\o-\frac32} }{2}} q_1^{\frac{1}{4}} q_1^{\frac{1}{2} (k^+_{\o+\frac12}-k^+_{\o+\frac32}+k_{2N-\o-\frac12}^--k^-_{2N-\o-\frac32}) } \right. \\
    & \qquad \times \left[ -e^{-b_{2N-\o-\frac{3}{2}} }q_1^{-\frac12} + (q_1^{-1}-1) \left( \bK_{+,\o+\frac{3}{2}} - \bK_{+,\o+\frac12} - \bK_{-,2N-\o-\frac12}^* + \bK_{-,2N-\o-\frac32}^* \right) \right] \\
    & \qquad \left. + \kq_{\o+\frac{1}{2}} e^{-\frac{b_{2N-\o-\frac12} }{2}} q_1^{\frac14} q_1^{\frac{1}{2}(k_{\o+\frac12}^+ - k_{\o-\frac12}^+ - k_{2N-\o+\frac12}^- + k_{2N-\o-\frac12}^- ) } \right \rangle 
\end{split}
\end{align}
DS equation: 
\begin{align}
\begin{split}
    0 & = \left \langle e^{\frac{b_{2N-\o-\frac32} }{2}} q_1^{\frac{1}{4}} q_1^{\frac{1}{2} (k^+_{\o+\frac12}-k^+_{\o+\frac32}+k_{2N-\o-\frac12}^--k^-_{2N-\o-\frac32}) } \right. \\
    & \qquad \times \left[ -e^{-b_{2N-\o-\frac{3}{2}} }q_1^{-\frac12} + (q_1^{-1}-1) \left( \bK_{+,\o+\frac{3}{2}} - \bK_{+,\o+\frac12} - \bK_{-,2N-\o-\frac12}^* + \bK_{-,2N-\o-\frac32}^* \right) \right] \\
    & \qquad + \kq_{\o+\frac{1}{2}} e^{-\frac{b_{2N-\o-\frac12} }{2}} q_1^{\frac14} q_1^{\frac{1}{2}(k_{\o+\frac12}^+ - k_{\o-\frac12}^+ - k_{2N-\o+\frac12}^- + k_{2N-\o-\frac12}^- ) } \\
    & \qquad + \left. {e^{-\frac{b_{2N-\o-\frac32}}{2}}q_1^{-\frac14}} q_1^{\frac{-1}{2}(k_{\o+\frac12}^+-k_{\o+\frac32}^++k_{2N-\o-\frac12}^- - k_{2N-\o-\frac32}^-)} \right\rangle
\end{split}
\end{align}
For $\o=0$, large $X$ expansion: 
\begin{align}
\begin{split}
    & \left[ X^{-\frac{1}{2}} \right] T_0(X) \\
    & = \left \langle e^{-\frac{b_{\frac32} }{2}} q_1^{\frac{1}{4}} q_1^{\frac{1}{2} (k^+_{\frac12}-k^+_{\frac32}+k_{-\frac12}^--k^-_{-\frac32}) } \right. \\
    & \qquad \left. \times \left[ -e^{b_{\frac{3}{2}} }q_1^{-\frac12} + (q_1^{-1}-1) \left( \bK_{+,\frac{3}{2}} - \bK_{+,\frac12} - \bK_{-,-\frac12}^* + \bK_{-,-\frac32}^* \right) \right] \right \rangle \\
    & \qquad + \kq_{\frac12} \left \langle e^{\frac{b_{\frac12}}{2}}q_1^{-\frac14} q_1^{\frac12 (k_{\frac12}^+-k_{-\frac12}^++k_{-\frac12}^--k_{\frac12}^-) } \right . \\
    & \qquad \left. \times \left[ e^{b_{\frac12}}q_1^{\frac12} + (q_1-1)q_2^{-1} \left( \bK_{+,\frac12} - \bK_{+,-\frac{1}{2}} + \bK_{-,-\frac12}^* - \bK_{-,\frac12}^* \right) \right] \right \rangle
\end{split}
\end{align}
DS equation: 
\begin{align}
\begin{split}
    0 & = \left \langle e^{-\frac{b_{\frac32} }{2}} q_1^{\frac{1}{4}} q_1^{\frac{1}{2} (k^+_{\frac12}-k^+_{\frac32}+k_{-\frac12}^--k^-_{-\frac32}) } \right. \\
    & \qquad \left. \times \left[ -e^{b_{\frac{3}{2}} }q_1^{-\frac12} + (q_1^{-1}-1) \left( \bK_{+,\frac{3}{2}} - \bK_{+,\frac12} - \bK_{-,-\frac12}^* + \bK_{-,-\frac32}^* \right) \right] \right \rangle \\
    & \qquad + \kq_{\frac12} \left \langle e^{\frac{b_{\frac12}}{2}}q_1^{-\frac14} q_1^{\frac12 (k_{\frac12}^+-k_{-\frac12}^++k_{-\frac12}^--k_{\frac12}^-) } \right . \\
    & \qquad \left. \times \left[ e^{b_{\frac12}}q_1^{\frac12} + (q_1-1)q_2^{-1} \left( \bK_{+,\frac12} - \bK_{+,-\frac{1}{2}} + \bK_{-,-\frac12}^* - \bK_{-,\frac12}^* \right) \right] \right \rangle \\
    & \qquad +  e^{\frac{b_\frac32 }{2}}q_1^{-\frac14} \left \langle q_1^{-\frac12 (k_{\frac12}^+-k_{\frac32}^++k_{-\frac12}^--k_{-\frac32}^-)} \right \rangle + \kq_{\frac{1}{2}} {e^{-\frac{b_{\frac12}}{2}}q_1^{\frac{1}{4}}} \left \langle q_1^{-\frac12 (k_{\frac12}^+-k_{-\frac12}^++k_{-\frac12}^--k_{\frac12}^-)} \right \rangle 
\end{split}
\end{align}
For $\o=N$, large $X$ expansion: 
\begin{align}
\begin{split}
    & \left[ X^{-\frac{1}{2}} \right] T_N(X) \\
    & = \left \langle e^{\frac{b_{N-\frac32} }{2}} q_1^{\frac{1}{4}} q_1^{\frac{1}{2} (k^+_{N+\frac12}-k^+_{N+\frac32}+k_{N-\frac12}^--k^-_{N-\frac32}) } \right. \\
    & \qquad \left. \times \left[ -e^{-b_{N-\frac{3}{2}} }q_1^{\frac12} + (q_1^{-1}-1) \left( \bK_{+,N+\frac{3}{2}} - \bK_{+,N+\frac12} - \bK_{-,N-\frac12}^* + \bK_{-,N-\frac32}^* \right) \right] \right \rangle \\
    & \qquad + \kq_{N+\frac12} \left \langle e^{\frac{-b_{N-\frac12}}{2}}q_1^{-\frac34} q_1^{\frac12 (k_{N+\frac12}^+-k_{N-\frac12}^++k_{N-\frac12}^--k_{N+\frac12}^-) } \right . \\
    & \qquad \left. \times \left[ e^{-b_{N-\frac12}}q_1^{-\frac12} + (q_1-1) \left( \bK_{+,N+\frac12} - \bK_{+,N-\frac{1}{2}} + \bK_{-,N-\frac12}^* - \bK_{-,N+\frac12}^* \right) \right] \right \rangle 
\end{split}
\end{align}
DS equation: 
\begin{align}
\begin{split}
    0 & = \left \langle e^{\frac{b_{N-\frac32} }{2}} q_1^{-\frac{1}{4}} q_1^{\frac{1}{2} (k^+_{N+\frac12}-k^+_{N+\frac32}+k_{N-\frac12}^--k^-_{N-\frac32}) } \right. \\
    & \qquad \left. \times \left[ -e^{-b_{N-\frac{3}{2}} }q_1^{\frac12} + (q_1^{-1}-1) \left( \bK_{+,N+\frac{3}{2}} - \bK_{+,N+\frac12} - \bK_{-,N-\frac12}^* + \bK_{-,N-\frac32}^* \right) \right] \right \rangle \\
    & \qquad + \kq_{N+\frac12} \left \langle e^{\frac{-b_{N-\frac12}}{2}}q_1^{-\frac34} q_1^{\frac12 (k_{N+\frac12}^+-k_{N-\frac12}^++k_{N-\frac12}^--k_{N+\frac12}^-) } \right . \\
    & \qquad \left. \times \left[ e^{-b_{N-\frac12}}q_1^{-\frac12} + (q_1-1) \left( \bK_{+,N+\frac12} - \bK_{+,N-\frac{1}{2}} + \bK_{-,N-\frac12}^* - \bK_{-,N+\frac12}^* \right) \right] \right \rangle \\
    & \quad + e^{\frac{-b_{N-\frac32}}{2}}q_1^{\frac14} \left \langle q_1^{-\frac12 (k_{N+\frac12}-k_{N+\frac32}+k_{N-\frac12}-k_{N-\frac32})} \right \rangle + \kq_{N+\frac12}  e^{\frac{b_{N-\frac12}}{2}}q_1^{\frac{3}{4}} \left \langle q_1^{-\frac12 (k_{N+\frac12}^+-k_{N-\frac12}^++k_{N-\frac12}^--k_{N+\frac12}^-)} \right \rangle
\end{split}
\end{align}
For $\o=N-1$, large $X$ expansion: 
\begin{align}
\begin{split}
    & -\left[ X^{-\frac{1}{2}} \right] T_{N-1}(X) \\
    & = \left \langle e^{\frac{b_{N-\frac12}}{2}}q^{\frac14} q_1^{\frac{1}{2}(k_{N-\frac12}^+-k_{N+\frac12}^+ + k^-_{N+\frac{1}{2}}-k^-_{N-\frac12}) } \right. \\
    & \qquad \times \left. \left[ - e^{-b_{N-\frac12}}q_1^{-\frac12} + (q_1^{-1}-1) \left( \bK_{+,N+\frac12} - \bK_{+,N-\frac12} - \bK_{-,N+\frac12}^* + \bK_{-,N-\frac12}^* \right) \right] \right \rangle \\
    & \quad - \kq_{N-\frac12} \left \langle  e^{\frac{b_{N-\frac12}}{2}} q_1^{\frac{3}{4}} q^{\frac12 (k^+_{N-\frac12}-k^+_{N-\frac32}+k^-_{N+\frac12}-k^-_{N+\frac32})} \right. \\
    & \qquad \times \left. \left[ e^{b_{N-\frac12}}q_1^{\frac12} + (q_1-1) \left( \bK_{+,N-\frac12} - \bK_{+,N-\frac32} + \bK_{-,N+\frac12}^* - \bK_{-,N+\frac32}^*  \right) \right] \right \rangle
\end{split}
\end{align}
DS equation: 
\begin{align}
\begin{split}
    0 & = \left \langle e^{\frac{b_{N-\frac12}}{2}}q^{\frac14} q_1^{\frac{1}{2}(k_{N-\frac12}^+-k_{N+\frac12}^+ + k^-_{N+\frac{1}{2}}-k^-_{N-\frac12}) } \right. \\
    & \qquad \times \left. \left[ - e^{-b_{N-\frac12}}q_1^{-\frac12} + (q_1^{-1}-1) \left( \bK_{+,N+\frac12} - \bK_{+,N-\frac12} - \bK_{-,N+\frac12}^* + \bK_{-,N-\frac12}^* \right) \right] \right \rangle \\
    & \quad - \kq_{N-\frac12} \left \langle  e^{\frac{b_{N-\frac12}}{2}q_1^{\frac{3}{4}}} q^{\frac12 (k^+_{N-\frac12}-k^+_{N-\frac32}+k^-_{N+\frac12}-k^-_{N+\frac32})} \right. \\
    & \qquad \times \left. \left[ e^{b_{N-\frac12}}q_1^{\frac12} + (q_1-1) \left( \bK_{+,N-\frac12} - \bK_{+,N-\frac32} + \bK_{-,N+\frac12}^* - \bK_{-,N+\frac32}^*  \right) \right] \right \rangle \\
    & \qquad + e^{\frac{b_{N-\frac12}}{2}}q_1^{-\frac14}\left\langle q_1^{-\frac{1}{2}(k_{N-\frac12}^+-k_{N+\frac12}^++k_{N+\frac12}^--k_{N-\frac12}^-)} \right\rangle
    - q_1^{-\frac34}\kq_{N-\frac{1}{2}} e^{-\frac{b_{N-\frac12}}{2}} \left \langle q_1^{-\frac{1}{2}(k_{N+\frac12}^-+k_{N-\frac12}^+-k_{N+\frac32}^--k_{N-\frac32}^+)} \right \rangle .
\end{split}
\end{align}
For $\o=2N-1$, large $X$ expansion: 
\begin{align}
\begin{split}
    & \left[ X^{-\frac{1}{2}} \right] T_{2N-1}(X) \\
    & = \left \langle e^{-\frac{b_{\frac12}}{2}}q_1^{\frac{1}{4}} q_1^{\frac{1}{2}(k_{-\frac12}^+-k_{\frac12}^++k_{\frac12}^--k_{-\frac12}^-)}  \right. \\
    & \qquad \times \left. \left[ - e^{b_{\frac12}}q_1^{-\frac12} q_2^{-1} + (q_1^{-1}-1) q_2^{-1} \left( \bK_{+,\frac12} - \bK_{+,-\frac12} + \bK_{-,-\frac12}^* - \bK_{-,\frac{1}{2}}^*  \right) \right] \right\rangle \\
    & \quad - \kq_{-\frac12} \left \langle e^{-\frac{b_{\frac12} }{2}} q_1^{\frac34} q_1^{\frac12 (k^+_{-\frac12}-k^+_{-\frac32} + k^-_{\frac12} - k^-_{\frac32} ) } \right. \\
    & \qquad \times \left. \left[ e^{-b_{\frac12}}q_1 + (q_1-1) \left( \bK_{+,-\frac12} - \bK_{+,-\frac32} + \bK_{-,\frac{1}{2}}^* - \bK_{-,\frac32}^* \right) \right] \right\rangle 
\end{split}
\end{align}
DS equation:
\begin{align}
\begin{split}
    0 & = \left \langle e^{-\frac{b_{\frac12}}{2}}q_1^{\frac{1}{4}} q_1^{\frac{1}{2}(k_{-\frac12}^+-k_{\frac12}^++k_{\frac12}^--k_{-\frac12}^-)}  \right. \\
    & \qquad \times \left. \left[ - e^{b_{\frac12}}q_1^{-\frac12} q_2^{-1} + (q_1^{-1}-1) q_2^{-1} \left( \bK_{+,\frac12} - \bK_{+,-\frac12} + \bK_{-,-\frac12}^* - \bK_{-,\frac{1}{2}}^*  \right) \right] \right\rangle \\
    & \quad - \kq_{-\frac12} \left \langle e^{-\frac{b_{\frac12} }{2}} q_1^{\frac34} q_1^{\frac12 (k^+_{-\frac12}-k^+_{-\frac32} + k^-_{\frac12} - k^-_{\frac32} ) } \right. \\
    & \qquad \times \left. \left[ e^{-b_{\frac12}}q_1 + (q_1-1) \left( \bK_{+,-\frac12} - \bK_{+,-\frac32} + \bK_{-,\frac{1}{2}}^* - \bK_{-,\frac32}^* \right) \right] \right\rangle \\
    & \quad + e^{\frac{b_{\frac12}}{2}}q_1^{-\frac14} \left \langle q_1^{-\frac{1}{2}(k_{-\frac12}^+-k_{\frac12}^++k_{\frac12}^--k_{-\frac12}^-)} \right\rangle - q_1^{-\frac{3}{4}} \kq_{-\frac{1}{2}}  e^{\frac{b_1}{2}}  \left\langle q_1^{-\frac{1}{2} (k_{\frac{1}{2}}+k_{-\frac{1}{2}}-k_{-\frac{3}{2}}-k_{\frac32}) } \right\rangle
\end{split}
\end{align}
We consider linear combination of the DS equations
\[
    \sum_{\o=0}^{2N-1} \hat{C}_\o \left( c^{(+)}_{1,\o} - c^{(-)}_{0,\o} \right)_\o = 0
\]
The coefficients are chosen to cancel unwanted terms $\bK_{\pm,\o+\frac12}$. 
\[
    0=\hat{C}_{2N-1} e^{-\frac{b_{\frac12}}{2}}q_1^{\frac14}q_1^{k^+_{-\frac12}-k^+_{\frac12}} - \hat{C}_0 \left( e^{-\frac{b_{\frac32}}{2}}q_1^{\frac14} q_1^{k^+_{\frac12} - k^+_{\frac32}} + \kq_{\frac12} e^{\frac{b_{\frac12}}{2}} q_1^{\frac{3}{4}}q_2^{-1} q_1^{k^+_{\frac12}-k^+_{-\frac12}} \right), \quad \o+{\frac12} = {\frac12}
\]
\[
    0=\hat{C}_{\o-1} e^{-\frac{b_{\o+\frac12}}{2}} q_1^{{\frac14}} q_1^{k^+_{\o-\frac12}-k^+_{\o+\frac12}} - \hat{C}_\o e^{-\frac{b_{\o+\frac32}}{2}} q_1^{\frac14} q_1^{k^+_{\o+\frac12}-k^+_{\o+\frac32}} ,\quad \o+{\frac12} = {\frac32},\dots,N-{\frac52}
\]
\[
    0=\hat{C}_{N-3} e^{-\frac{b_{N-\frac32}}{2}} q_1^{{\frac14}} q_1^{k^+_{N-\frac52}-k^+_{N-\frac32}} - \hat{C}_{N-2} e^{-\frac{b_{N-\frac12}}{2}} q_1^{\frac14} q_1^{k^+_{N-\frac32}-k^+_{N-\frac12}} + \hat{C}_{N-1} \kq_{N-\frac12} e^{\frac{b_{N-\frac12}}{2}} q_1^{\frac{7}{4}} q_1^{k^+_{N-\frac12}-k^+_{N-\frac32}}, \quad \o+{\frac12} = N-{\frac32}
\]
\begin{align}
\begin{split} \nonumber
    0= & \ \hat{C}_{N-2} e^{-\frac{b_{N-\frac12}}{2}} q_1^{\frac14} q_1^{k^+_{N-\frac32}-k^+_{N-\frac12}}  - \hat{C}_{N-1} \left( e^{\frac{b_{N-\frac12}}{2}} q_1^{\frac14} q_1^{k^+_{N-\frac12}-k^+_{N+\frac12}} + \kq_{N-\frac12} e^{\frac{b_{N-\frac12}}{2}} q_1^{\frac74} q_1^{k^+_{N-\frac12}-k^+_{N-\frac32}} \right) \\
    & + \hat{C}_N \kq_{N+\frac12}e^{-\frac{b_{N-\frac12}}{2}} q_1^{\frac14} q_1^{k^+_{N+\frac12}-k^+_{N-\frac12}} ,\quad \o+{\frac12} = N-{\frac12}
\end{split}
\end{align}
\[
    0=\hat{C}_{N-1} e^{\frac{b_{N-\frac12}}{2}} q_1^{\frac14} q_1^{k^+_{N-\frac12}-k^+_{N+\frac12}} - \hat{C}_N \left( e^{\frac{b_{N-\frac32}}{2}} q_1^{\frac14} q_1^{k^+_{N+\frac12}-k^+_{N+\frac32}} + \kq_{N+\frac12} e^{-\frac{b_{N-\frac12}}{2}} q_1^{\frac14} q_1^{k^+_{N+\frac12}-k^+_{N-\frac12}}  \right), \quad \o+\frac12 = N+\frac12
\]
\[
    0=\hat{C}_{\o-1} e^{\frac{b_{2N-\o-\frac12}}{2}} q_1^{\frac14} q_1^{k^+_{\o-\frac12}-k^+_{\o+\frac12}} - \hat{C}_\o e^{\frac{b_{2N-\o-\frac32}}{2}} q_1^{\frac14} q_1^{k^+_{\o-\frac12}-k^+_{\o+\frac32}} ,\quad \o+\frac12 = N+\frac32,\dots,2N-\frac52
\]
\[
    0=\hat{C}_{2N-3} e^{\frac{b_{\frac32}}{2}} q_1^{\frac14} q_1^{k^+_{-\frac52}-k^+_{-\frac32}} -\hat{C}_{2N-2} e^{\frac{b_{\frac12}}{2}} q_1^{\frac14} q_1^{k^+_{-\frac32}-k^+_{-\frac12}} - \hat{C}_{2N-1} \kq_{-\frac12} e^{-\frac{b_{\frac12}}{2}} q_1^{\frac74} q_1^{k^+_{-\frac12}-k^+_{-\frac32}} ,\quad \o + \frac12 = 2N - \frac32 
\]
\begin{align}
\begin{split} \nonumber
    0=& \ \hat{C}_{2N-2} e^{\frac{b_{\frac12}}{2}} q_1^{\frac14} q_1^{k^+_{\frac32}-k^+_{-\frac12}} - \hat{C}_{2N-1} \left( e^{-\frac{b_{\frac12}}{2}}q_1^{\frac14} q_1^{k^+_{-\frac12}-k^+_{\frac12}} q_2^{-1} - \kq_{-\frac12} e^{-\frac{b_{\frac12}}{2}} q_1^{\frac74} q_1^{k^+_{-\frac12}-k^+_{-\frac32}} \right) \\
    & - \hat{C}_0 \kq_{\frac12} e^{\frac{b_{\frac12 }}{2}} q_1^{\frac34} q_2^{-1} q_1^{k^+_{\frac12}-k^+_{-\frac12}} ,\quad \o+\frac12 = 2N - \frac12
\end{split}
\end{align}
We find the following solution: 
\[
    \hat{C}_\o = e^{\frac{b_{\o+\frac32} }{2}}q_1^{-\frac14} q_1^{-\nabla^{u}_{\o+\frac32} } 
    \ \o = 0,\dots,N-3
\]
\[
    \hat{C}_\o = e^{-\frac{b_{2N-\o-\frac32}}{2}} q_1^{-\frac14} q_1^{-\nabla^u_{\o+\frac32}}
    \ \o = N,\dots, 2N-3
\]
\[
    \hat{C}_{N-1} = e^{-\frac{b_{N-\frac12}}{2}} q_1^{-\frac14} q_1^{-\nabla^u_{N+\frac12}}
    + e^{-b_{N-\frac12} - \frac{b_{N-\frac32}}{2} } q_1^{-\frac14} 
    q_1^{-\nabla^u_{N+\frac12}-\nabla^u_{N+\frac32} } 
    \kq_{N+\frac12} 
    q_1^{-\nabla^u_{N+\frac12}} 
\]
\begin{align}
\begin{split} \nonumber
    \hat{C}_{N-2} = & \ e^{\frac{b_{N-\frac12}}{2}} q_1^{-\frac14} 
    q_1^{-\nabla^u_{N-\frac12} } 
    + e^{\frac{b_{N-\frac12} }{2}} q_1^{\frac54} 
    q_1^{-\nabla^u_{N+\frac12}-\nabla^u_{N-\frac12}} 
    \kq_{N-\frac12} 
    q_1^{-\nabla^u_{N-\frac12}} 
    \\
    & + e^{-\frac{b_{N-\frac32}}{2}} q_1 
    q_1^{-\nabla^u_{N+\frac32} - \nabla^u_{N+\frac12} - \nabla^u_{N-\frac12} } 
    \kq_{N+\frac12} 
    q_1^{-\nabla^u_{N+\frac12}} 
    \kq_{N-\frac12} 
    q_1^{-\nabla^u_{N-\frac12}} 
\end{split}
\end{align}
\[
    \hat{C}_{2N-1} = e^{\frac{b_{\frac12}}{2}} q_1^{-\frac14} q_1^{-\nabla^u_{\frac12}} + e^{b_{\frac12}+\frac{b_{\frac32}}{2}} q_1^{\frac14} q_2^{-1} q_1^{-\nabla^u_\frac12 - \nabla^u_{\frac32}} \kq_\frac12 q_1^{-\nabla^u_{\frac12}}
\]
\begin{align}
\begin{split} \nonumber
    \hat{C}_{2N-2}
    = & \ e^{-\frac{b_\frac12}{2}} q_1^{-\frac14} q_1^{-\nabla^u_{2N-\frac12}} - e^{-\frac{b_\frac12}{2}} q_1^{\frac54} q_1^{-\nabla^u_{2N-\frac12}-\nabla^u_{\frac12}} \kq_{2N-\frac12} q_1^{-\nabla^u_{2N-\frac12}} \\
    & - e^{\frac{b_{\frac32}}{2}} q_1^{\frac{7}{4}} q_1^{-\nabla^u_{2N-\frac12} - \nabla^u_{\frac12} - \nabla^u_{\frac32} } \kq_{\frac12} q_1^{-\nabla^u_{\frac12}} \kq_{2N-\frac12} q_1^{-\nabla^u_{2N-\frac12}}
\end{split}
\end{align}

The summation over Dyson-Schwinger equation can be written as differential operators acting on the defect partition function
\begin{align}
\begin{split}
    0 = \sum_{\o=0}^{2N-1} \hat{C}_{\o} (\text{DS})_\o = \left \langle \left( \hat{\rm H} - \bS - \bS^* \right) \Psi \right \rangle. 
\end{split}
\end{align}
The Hamiltonian is given by
\begin{align}
\begin{split}
    \hat{\rm H} = &\sum_{\o=0}^{2N-1} e^{-\hat{p}_{\o+\frac12}} + \sum_{\o=2}^{N-2} 2 \kq_{\o-\frac12} e^{-\frac{\hat{p}_{\o+\frac12}+\hat{p}_{\o-\frac12}}{2}} + \sum_{\o=N+2}^{2N-2} 2 \kq_{\o-\frac12} e^{-\frac{\hat{p}_{\o+\frac12}+\hat{p}_{\o-\frac12}}{2}} \\
    & + \kq_{N+\frac12}\kq_{N-\frac12}\kq_{N-\frac32} e^{\frac{-\hat{p}_{N+\frac32}-2\hat{p}_{N+\frac12}-2\hat{p}_{N-\frac12}-\hat{p}_{N-\frac32}}{2}} + \kq_{N-\frac12} \left( e^{\frac{-\hat{p}_{N+\frac12}-3\hat{p}_{N-\frac12}}{2}} + e^{\frac{\hat{p}_{N-\frac12}-\hat{p}_{N+\frac12}}{2}} \right) \\
    & + \kq_{N-\frac{3}{2}} \left(1+\kq_{N-\frac12} e^{\frac{-\hat{p}_{N-\frac12}-\hat{p}_{N+\frac12}}{2}}\right) e^{\frac{\hat{p}_{N-\frac12}-\hat{p}_{N-\frac32}}{2}} \\
    & + \kq_{N+\frac12} \left( e^{\frac{\hat{p}_{N+\frac12}-\hat{p}_{N+\frac32}}{2}} + e^{\frac{-3\hat{p}_{N+\frac12}-\hat{p}_{N+\frac32}}{2}} \right) + \kq_{N+\frac12} \kq_{N-\frac12} \left( e^{\frac{\hat{p}_{N-\frac12}-2\hat{p}_{N+\frac12}-\hat{p}_{N+\frac32}}{2}} + e^{\frac{-3\hat{p}_{N-\frac12}-2\hat{p}_{N+\frac12}-\hat{p}_{N+\frac32}}{2}} \right) \\
    & + \kq_{\frac12}\kq_{2N-\frac12}\kq_{2N-\frac32} e^{\frac{-\hat{p}_{\frac32}-2\hat{p}_{\frac12}-2\hat{p}_{2N-\frac12}-\hat{p}_{2N-\frac32}}{2}} + \kq_{2N-\frac12} \left( e^{\frac{-\hat{p}_{\frac12}-3\hat{p}_{2N-\frac12}}{2}} + e^{\frac{\hat{p}_{2N-\frac12}-\hat{p}_{\frac12}}{2}} \right) \\
    & + \kq_{2N-\frac{3}{2}} \left(1+\kq_{2N-\frac12} e^{\frac{-\hat{p}_{2N-\frac12}-\hat{p}_{\frac12}}{2}}\right) e^{\frac{\hat{p}_{2N-\frac12}-\hat{p}_{2N-\frac32}}{2}} \\
    & + \kq_{\frac12} \left( e^{\frac{\hat{p}_{\frac12}-\hat{p}_{\frac32}}{2}} + e^{\frac{-3\hat{p}_{\frac12}-\hat{p}_{\frac32}}{2}} \right) + \kq_{\frac12} \kq_{2N-\frac12} \left( e^{\frac{\hat{p}_{2N-\frac12}-2\hat{p}_{\frac12}-\hat{p}_{\frac32}}{2}} + e^{\frac{-3\hat{p}_{2N-\frac12}-2\hat{p}_{2N+\frac12}-\hat{p}_{2N+\frac32}}{2}} \right) \\
\end{split}
\end{align}
with $\hat{p}_{\o+\frac{1}{2}} = 2\ve_1  \nabla^u_{{\o+\frac12}} $.

\bibliographystyle{utphys}
\bibliography{SO-SU}

\end{document}